\documentclass{SciPost}

\usepackage{siunitx}
\usepackage{hyperref}
\usepackage{orcidlink}
\usepackage[font={small}]{caption}
\usepackage{subcaption}
\usepackage{float}
\usepackage{multirow}
\usepackage{xspace}
\usepackage{makecell,booktabs}
\usepackage[normalem]{ulem}

\hypersetup{
    colorlinks,
    linkcolor={red!50!black},
    citecolor={blue!50!black},
    urlcolor={blue!80!black}
}

\usepackage[bitstream-charter]{mathdesign}
\DeclareSymbolFont{usualmathcal}{OMS}{cmsy}{m}{n}
\DeclareSymbolFontAlphabet{\mathcal}{usualmathcal}

\newcommand{\rivet}{\textsc{Rivet}\xspace}
\newcommand{\yoda}{\textsc{Yoda}\xspace}
\newcommand{\hepdata}{\textsc{HEPData}\xspace}
\newcommand{\contur}{\textsc{Contur}\xspace}

\newcommand{\madgraph}{\textsc{MadGraph5\_aMC@NLO}\xspace}
\newcommand{\powheg}{\textsc{Powheg}\xspace}
\newcommand{\pepper}{\textsc{Pepper}\xspace}
\newcommand{\sherpa}{\textsc{Sherpa}\xspace}
\newcommand{\pythia}{\textsc{Pythia}\xspace}
\newcommand{\herwig}{\textsc{Herwig}\xspace}
\newcommand{\delphes}{\textsc{Delphes}~3\xspace}
\newcommand{\hepmc}{\textsc{HepMC}\xspace}
\newcommand{\checkmate}{\textsc{CheckMATE}\xspace}
\newcommand{\madanalysis}{\textsc{MadAnalysis}~5\xspace}
\newcommand{\xrootd}{\textsc{xRootD}\xspace}
\newcommand{\miniaod}{\textsc{MiniAOD}\xspace}
\newcommand{\nanoaod}{\textsc{NanoAOD}\xspace}
\newcommand{\miniaodsim}{\textsc{MiniAODSIM}\xspace}
\newcommand{\nanoaodsim}{\textsc{NanoAODSIM}\xspace}
\newcommand{\lhe}{LHE\xspace}
\newcommand{\lhehfive}{LHEH5\xspace}
\newcommand{\hdffive}{HDF5\xspace}

\fancypagestyle{SPstyle}{
\fancyhf{}
\lhead{\colorbox{scipostblue}{\bf \color{white} ~SciPost Physics Community Reports }}
\rhead{{\bf \color{scipostdeepblue} ~Submission }}

\fancyfoot[C]{\textbf{\thepage}}
}

\begin{document}

\pagestyle{SPstyle}

%%%%%%%%%% paper title %%%%%%%%%%%%%%%%%%%%%%%%%%%%%%%%%%
\begin{center}{\Large \textbf{\color{scipostdeepblue}{
Open LHC Monte Carlo Event Generation\\
}}}\end{center}
%%%%%%%%%%%%%%%%%%%%%%%%%%%%%%%%%%%%%%%%%%%%%%%%%%%%%%%%%

%%%%%%%%%% AUTHORS 
\begin{center}\textbf{
Enrico Bothmann\,\orcidlink{0000-0001-6786-6843}\textsuperscript{1},
Jon Butterworth\,\orcidlink{0000-0002-5905-5394}\textsuperscript{2},
Shu Chen\,\orcidlink{}\textsuperscript{3},
Eda Erdogan\,\orcidlink{}\textsuperscript{1,4},
Giovanni~Guerrieri\,\orcidlink{0000-0002-3403-1177}\textsuperscript{1},
Christian~G\"utschow\,\orcidlink{0000-0003-0857-794X}\textsuperscript{5},
Martin Habedank\,\orcidlink{0000-0002-7412-9355}\textsuperscript{6},
Julie~M.~Hogan\,\orcidlink{0000-0002-8604-3452}\textsuperscript{7$\star$},
Venus~Keus\,\orcidlink{0000-0002-0345-3414}\textsuperscript{1,8,9},
Sabine~Kraml\,\orcidlink{0000-0002-2613-7000}\textsuperscript{10$\dagger$},
Van~Dung~Le\,\orcidlink{https://orcid.org/0009-0006-2306-1000}\textsuperscript{9,11}, 
Kati~Lassila-Perini\,\orcidlink{0000-0002-5502-1795}\textsuperscript{9}, 
Rakhi~Mahbubani\,\orcidlink{0000-0001-6311-4310}\textsuperscript{12$\ddag$},
Zach~Marshall\,\orcidlink{0000-0003-0786-2570}\textsuperscript{13$\S$},
Thomas~McCauley\,\orcidlink{0000-0001-6589-8286}\textsuperscript{14},
Tomasz~Procter\,\orcidlink{0000-0002-6534-9153}\textsuperscript{15},
Humberto~Reyes-González\,\orcidlink{https://orcid.org/0000-0003-3283-5208}\textsuperscript{16},
Andrzej~Siodmok\,\orcidlink{0000-0001-9614-7856}\textsuperscript{15},
Mariana~Vivas~Albornoz\,\orcidlink{0009-0000-1453-5346}\textsuperscript{17} \\
(LHC REI WG Open Event Generation Task Force)\\
  }\end{center}

%%%%%%%%%% AFFILIATIONS
\begin{center}
%Enrico, Giovanni
{\bf 1} CERN, European Laboratory for Particle Physics, Geneva, Switzerland \\
%Jon
{\bf 2} Department of Physics and Astronomy, University College London, London, UK \\
%Shu
{\bf 3} School of Physics and Astronomy, University of Southampton, Southampton, UK\\
%Eda
{\bf 4} Yıldız Technical University, Istanbul, Turkiye \\
%Chris
{\bf 5} Centre for Advanced Research Computing, University College London, London, UK \\
%Martin
{\bf 6} School of Physics and Astronomy, University of Glasgow, Glasgow, UK \\
%Julie
{\bf 7} Bethel University, Saint Paul, MN, USA \\
%Venus
{\bf 8} School of Theoretical Physics, Dublin Institute for Advanced Studies, Dublin, Ireland \\
{\bf 9} Helsinki Institute of Physics, Helsinki, Finland \\
%Sabine
{\bf 10} Univ. Grenoble Alpes, CNRS, Grenoble INP, LPSC-IN2P3, Grenoble, France \\
%Van (+Helsinki IoP)
{\bf 11} Lappeenranta University of Technology, Lappeenranta, Finland\\
%Rakhi
{\bf 12} Rudjer Boskovic Institute, Division of Theoretical Physics, Zagreb, Croatia \\
%Zach
{\bf 13} Physics Division, Lawrence Berkeley National Laboratory, Berkeley, CA, USA \\
%Thomas
{\bf 14} University of Notre Dame, Notre Dame, IN, USA\\
%Tomek & Andrzej
{\bf 15} {Jagiellonian University, Krak\'{o}w, Poland}\\
%Humberto
{\bf 16} Institute for Theoretical Particle Physics and Cosmology, RWTH Aachen University, Aachen, Germany \\
%Mariana
{\bf 17} Deutsches Elektronen-Synchrotron DESY, Hamburg, Germany  \\[\baselineskip]

%%%%%%%%%% editors, corresponding authors
Corresponding authors: \\
$\star$~\href{mailto:j.hogan@cern.ch}{\small j.hogan@cern.ch}\,,~ 
$\dagger$~\href{mailto:sabine.kraml@lpsc.in2p3.fr}{\small sabine.kraml@lpsc.in2p3.fr}\,,\\
$\ddag$~\href{mailto:rakhi.mahbubani@cern.ch}{\small rakhi@irb.hr}\,,~ 
$\S$~\href{mailto:zach.marshall@cern.ch}{\small zach.marshall@cern.ch}\\
\end{center}

%%%%%%%%%% ABSTRACT %%%%%%%%%%
\section*{\color{scipostdeepblue}{Abstract}}
\textbf{\boldmath{%
The LHC physics programme involves a vast amount of Monte Carlo event simulation. This paper reviews current efforts towards sharing the generated events as Open Data. Open Event Generation helps reduce duplication of effort and resource consumption, and benefits the whole High Energy Physics community. 
We give examples of use cases and user experiences, discuss financial and environmental savings, and suggest future directions.   
}}

\vspace{\baselineskip}

%%%%%%%%%% BLOCK: Copyright information
% This block will be filled during the proof stage, and finilized just before publication.
% It exists here only as a placeholder, and should not be modified by authors.
\noindent\textcolor{white!90!black}{%
\fbox{\parbox{0.975\linewidth}{%
\textcolor{white!40!black}{\begin{tabular}{lr}%
  \begin{minipage}{0.6\textwidth}%
    {\small Copyright attribution to authors. \newline
    This work is a submission to SciPost Phys. Comm. Rep. \newline
    License information to appear upon publication. \newline
    Publication information to appear upon publication.}
  \end{minipage} & \begin{minipage}{0.4\textwidth}
    {\small Received Date \newline Accepted Date \newline Published Date}%
  \end{minipage}
\end{tabular}}
}}
}

% For convenience during refereeing we turn on line numbers:
%\linenumbers

%%%%%%%%%% TOC %%%%%%%%%%
\clearpage
\setcounter{tocdepth}{2}
\vspace{10pt}
\noindent\rule{\textwidth}{1pt}
\tableofcontents
\noindent\rule{\textwidth}{1pt}
\vspace{10pt}

%%%%%%%%%%%%%%%%%%%%%%%%%%%%%%%%%%%%%%%%%%%%%%%%%%%%%%%%%%%%%%%%%%%%%%%%%%%%%%%%
\section{Introduction}
\label{sec:intro}
%%%%%%%%%%%%%%%%%%%%%%%%%%%%%%%%%%%%%%%%%%%%%%%%%%%%%%%%%%%%%%%%%%%%%%%%%%%%%%%%

Physics studies at the Large Hadron Collider (LHC) require intensive Monte Carlo (MC) event simulation. 
For the experiments, this includes the simulation of large Standard Model (SM) and beyond-the-SM (BSM) event samples to design analyses, develop triggers and taggers, determine sensitivities, compare models against data, and so on. 
Collider phenomenologists do much the same, albeit outside the large experimental collaborations. For event generation, both communities typically use the same set of publicly available tools. There is thus significant duplication of effort and resource consumption, which can be mitigated by the sharing of simulated events across the High Energy Physics (HEP) community.

For this reason a recent white paper \cite{LHCReinterpretationForum:2025zgq}, submitted to the European Strategy for Particle Physics Update 2026,  recommended that 
\emph{``centralising the production of SM event samples to a core LHC team working in consultation with Monte Carlo experts should
be prioritised as a strategic objective, resulting in substantial financial and environmental savings and collateral benefits in increased robustness, transparency, and equitable access.''} 
The same white paper also advocated an effective curation of these data, and their release in a format that is compatible with
a large set of downstream phenomenological tools. 

Within HEP, open sharing of simulated data was pioneered by the Lattice community in the early 2000s, in the form of the International Lattice Data Grid (ILDG), an initiative to share resource-intensive simulations of gauge configurations outside the collaborations that generated them~\cite{SimmaTalk2023}, in line with FAIR principles~\cite{FAIR:2016}.  Although the first version of the effort became unusable due to lack of funding, the idea was recently revived~\cite{Karsch:2022tqw} and community efforts to revive ILDG 2.0 are well under way.  In the Cosmology community, the output of large cosmological simulations are stored within the Virgo database~\cite{VirgoDB}, which provides an online interface for user queries to an SQL database.

Both the ATLAS and the CMS collaborations have made large datasets of simulated events available through the CERN Open Data Portal~\cite{CODP}.\footnote{To our knowledge, no event generation data has been released by ALICE or LHCb.} CMS has been doing so since 2014 using the AODSIM format, and later the \miniaodsim and \nanoaodsim formats~\cite{Rizzi:2019rsi}. In both collaborations the releases include both SM and BSM processes that are simulated using a variety of common MC event generators. This ``event generation data'' is then processed through a simulation of the detector apparatus, and finally the experiment's reconstruction algorithms can be applied to the simulated event. Experimental simulations released as Open Data thus contain event generation information as well as ``reconstructed'' information. Following a community request~\cite{Marshall2025}, ATLAS extended its Open Data offerings with a first release of generated events in \hepmc format~\cite{hepmc} in 2025.   

It is thus timely to discuss Open LHC Event Generation in more detail and with a wider perspective. This is the purpose of this paper.  
Concretely, we first review in Section~\ref{sec:available-evtgen} the current ATLAS and CMS practices regarding the release of generator-level data.  
In Section~\ref{sec:usecases} we give examples of use cases, including user experiences that might serve as feedback for future releases. The financial and environmental benefits of making simulated events publicly available are discussed in Section~\ref{sec:sustainability}.
Indeed, event generation for the LHC places heavy demands on computing infrastructure, electricity and CPU time, and human resources, resulting in significant financial cost.   
Event generation also has an environmental cost, not only from
generating the electricity required to run the simulations, but also in the form of ``embodied carbon'', the emissions due to manufacturing the computing infrastructure on which the codes
run, and disposing of it once it reaches the end of its useful lifespan.  
Future directions are laid out in Section~\ref{sec:future}. 
We conclude in Section~\ref{sec:conclusion} with a summary and a list of concrete action items for the short, medium and long term. 

Our objectives are to raise awareness of the issues discussed here, increase the utility and uptake of the data that is already provided, motivate future releases and Open Data infrastructure improvements, and finally drive efforts to centralise MC generation to a core LHC team as an integral part of future European Strategy~\cite{LHCReinterpretationForum:2025zgq}.

%%%%%%%%%%%%%%%%%%%%%%%%%%%%%%%%%%%%%%%%%%%%%%%%%%%%%%%%%%%%%%%%%%%%%%%%%%%%%%%%
\section{Available Open Event Generation}
\label{sec:available-evtgen}
%%%%%%%%%%%%%%%%%%%%%%%%%%%%%%%%%%%%%%%%%%%%%%%%%%%%%%%%%%%%%%%%%%%%%%%%%%%%%%%%

The LHC experiments have Open Data programs through which they release data and MC simulation datasets to the community. The datasets are available as records on the CERN Open Data Portal~\cite{CODP} and can be queried using the \verb|cernopendata-client| command-line tool~\cite{cernopendataclient}. All data are released under a CC0 license~\cite{cc0} and may be freely used with appropriate attribution of the dataset's DOI~\cite{citingatlas,citingcms}. 
The releases have historically focused on collision data and simulation in data formats used by the experiments themselves for analysis. 
While these data offer substantial value, certain research directions require event generator information in addition, or instead.
Likewise, not all use-cases can be satisfied by Open Event Generation data; some limitations are mentioned in Section~\ref{sec:usecases}.
The experiments also release significant information via the \hepdata repository~\cite{HEPData}. This is generally limited to digitised figures and metadata, and does not generally contain any event-wise data. 
Although of great value to the community, these (usually analysis-specific) data products~\cite{Bailey:2022tdz} cannot replace event-level data, real or simulated, in any meaningful way.

This section describes the current status of ATLAS and CMS Open Data related to event generation, including dedicated event generation releases and Monte Carlo datasets containing generator level information.
Feedback on the information presented, and on the open data itself, is welcome through the CERN Open Data discussion forum~\cite{od-forum}.

%%%%%%%%%%%%%%%%%%%%%%%%%%%%%%%%%%%%%%%%%%%%%%%%%
\subsection {ATLAS Open Event Generation data}
%%%%%%%%%%%%%%%%%%%%%%%%%%%%%%%%%%%%%%%%%%%%%%%%%
\label{sec:atlas_evgen}

In 2024 the ATLAS collaboration released simulated and collision data collected during 2015--2016 at $\sqrt{s}=13$~TeV in the ATLAS ``PHYSLITE'' format~\cite{atlasopedata}. In addition to reconstructed objects, this data format includes derived quantities, such as collections of electrons and particle jets, but contains only limited generator information due to restrictions on file size. 
Acting on a community request, ATLAS subsequently released dedicated event generation data in \hepmc-format~\cite{hepmc}, comprised of 12.8~billion generated events in 6,213 datasets, totalling around 900~TB of data~\cite{atlasopenevngen}. Samples are available at 13~TeV (Run 2-like) and 13.6~TeV (Run 3-like) centre-of-mass energies. In general, samples contain the equivalent of double the integrated luminosity available at that centre-of-mass energy (e.g.~about 280~fb$^{-1}$ equivalent at 13~TeV), with no fewer than 10,000 events (in order to provide reasonable statistics for distributions), and no more than 10~million events (in order to save storage resources). In the following sections we focus on these samples as an example of the public release of dedicated event generation data in a standard format.

\subsubsection{Data format}
\label{sec:atlas_dataformat}

The data are released in tarred and gzipped \hepmc-format text files, currently \hepmc version 2. The datasets generally contain files with 10,000 events each, and users are free to include as many of these files from each dataset as they wish. The production of these samples from ATLAS internal generated event datasets proceeds through a thin conversion layer: a single file in the internal ``EVNT'' format (a ROOT-based format that includes both the \hepmc event and a container storing event metadata) is converted into the text \hepmc output format. 

The conversion runs at about 25~events/s, with gzip compression accounting for roughly two-thirds of the runtime. This compression is beneficial: compressed \hepmc is similar in size to the EVNT format ATLAS uses, while uncompressed \hepmc is about four times larger. Tests performed after the initial data release demonstrated that the tarring process did not save any additional space relative to gzip. Therefore, future samples will be released in a gzipped text format, which has the advantage that programs can directly read the data without needing to decompress the files first. 
Only minor modifications to the event record as written by the event generator are made internally, before the event records are stored. For example, an event weight is added if none is available, %particles with a PDG identifier of 0 are removed, 
a signal vertex is added in case none is set by the generator and loops written by the event generator are generally removed.\footnote{Loops in event records, which are particularly common in some event generators like \sherpa~\cite{Sherpa:2019gpd,Sherpa:2024mfk}, are problematic for any software that attempts to traverse the event record like it is a tree. Depending on the output structure the event generator writes, they can occur in higher-order matrix elements and in some hadronisation and shower configurations like the dipole shower used in Sherpa. Additionally, because of the way \hepmc library version~3 \cite{Buckley:2019xhk} keeps event data in memory, they can induce crashes when, for example, retrieving the position of a vertex.} 

ATLAS also produces ROOT-based ``derivation'' formats with \SI{30}{\percent} smaller size, which allow, in addition, the inclusion of analysis-friendly collections (e.g.~a collection of pre-selected electrons or jets) for about \SI{10}{\percent} additional disk space. However, these derivation formats cannot currently be read by the tools the phenomenology community uses, so they were disfavoured for this Open Data release. \hepmc version~3 format was also considered for output, but it currently requires about \SI{40}{\percent} more disk space than \hepmc version~2 format after compression. This is due primarily to the use of relatively inefficient storage for ``attributes'' in the \hepmc version~3 format, which are required to reproduce the full contents of the \hepmc version~2 format contents. For example, barcodes are not a part of the version~3 format, and so must be stored as attributes if they are to be retained. The \hepmc version~3 format also has higher precision for floating point numbers by default: version~2 records 32-bit floating point numbers, while version~3 records more than 64 bits of precision by default. Weight names do not appear in either format natively. Investigations how to adapt the output format to achieve small enough file size without loss of relevant information are ongoing. 

\subsubsection{Data curation and availability}

The event generator data are transferred to a Rucio~\cite{Barisits_2019} endpoint at CERN, hosted on a public file system (i.e.,~not on ATLAS resources). From there they can be read using the https~\cite{https} or \xrootd~\cite{xrootd} protocols, or read directly by systems with access to CERN file systems like lxplus~\cite{lxplus}, SWAN~\cite{swan19}, or the ESCAPE VRE~\cite{escapevre}. The files are also available to Rucio users within ATLAS, and with upcoming developments they should, in addition, be available to Rucio users outside of ATLAS. The samples are gathered into records for the CERN Open Data Portal, with around 80 records containing the samples. Each CERN Open Data Portal record contains a set of like samples: all signals related to a specific new physics model are collected together, for example, as are all Standard Model samples including a Higgs boson and top quarks. The division among Open Data Portal records is done using the keywords assigned to each sample. Two ``entry point'' records are available, linking all samples with the same centre-of-mass energy. These are meant to provide easier identification of appropriate samples, as well as to provide a single high-level DOI for citation (rather than requiring the citation of a DOI for each individual sample).

The datasets contain all standard ``nominal'' samples used within ATLAS for all major Standard Model physics processes; samples used for standard systematic variations (but not including specialised variations like different top masses); and samples often used as alternative nominal samples for analyses with special requirements (e.g.~a top-quark sample with a better description of many-jet final states). The combination of appropriate samples to create a complete background estimate is non-trivial. For example, several samples of QCD jet production must be combined to recreate the full physical spectrum; a single sample represents only a range in transverse momentum. Documentation has been prepared indicating the expected nominal Standard Model samples that should be combined, as well as indicating those that should be reserved for studies of systematic variations. The release also includes about 2,000 samples of new physics across a variety of models, including new vector bosons, leptoquarks, dark sectors, heavy neutral leptons, and supersymmetric models. 

\subsubsection{Metadata}

For each sample, metadata are provided, including:

\begin{itemize}
    \item the generators and their versions used, as well as the version and configuration of ATLAS software used to produce the sample;
    \item the sample cross section (as calculated by the event generator) with uncertainty when it is available, filters used and the filter efficiency (in case events have been filtered, for example by selecting only events with leptons), and a $k$-factor for samples that can be normalised to a higher-order cross section calculation;
    \item the centre-of-mass energy of the sample;
    \item the ``physics short'' of the sample, which is a 50--60 character name used to uniquely identify and describe the physics configuration of each sample;
    \item the number of events available publicly, the \hepmc format version (currently always ``2''), and the number of events available internally, so that users are able to quickly identify samples for which additional events could be made available upon well-motivated request;
    \item the tune of non-perturbative parameters used in generation, as well as the parton distribution function (PDF) used for the sample;
    \item a human-readable process description;
    \item a list of keywords identifying the categories in which the samples belong, where the keywords include physics (e.g.~``Higgs''), generation (e.g.~``NLO''), final state (e.g.~``1lepton''), and usage (e.g.~``performance'') indicators;
    \item a link to the ``job options'', a python file in which the precise generator configuration is specified. These files are written for configuration of the ATLAS software, but often the generator configuration can be straightforwardly extracted.
\end{itemize}

These metadata are made available through the \texttt{atlasopenmagic} python package~\cite{atom25}, which provides easy access to files and their locations as well as the metadata for the datasets. \texttt{atlasopenmagic} also includes convenient search functions to identify samples based on their metadata (e.g.~keywords, descriptions, or cross sections). It can provide file locations based on the preferred protocol, and in case a user has downloaded a significant volume of data it can identify known datasets on the local file system.

The metadata available for each sample are not always complete. They are extracted from the ATLAS internal metadata database (AMI), tools provided by the ATLAS Physics Modelling Group, and from the job options files themselves, and in some cases additions are made in order to ensure appropriate keywords for the samples. The effort to improve these metadata is ongoing.

\subsubsection{Usage}

To demonstrate the usage of the open event generation data, a short notebook~\cite{ATLAStutorial} has been prepared that explains how to search the open event generation samples, identify the samples of interest, and get files; how to inspect files from the samples and understand the event generator records using \texttt{pyhepmc}~\cite{pyhepmc}; how to make simple histograms from the input files using \texttt{hist}~\cite{hist}; and how to run a simple parametric simulation (e.g.~\delphes~\cite{deFavereau:2013fsa}) 
using the files as input and examine the simulated output using \texttt{uproot}~\cite{uproot}. A tutorial is also available describing how to make new samples that are consistent with those from the ATLAS Open Data, although the tutorial has not yet been thoroughly vetted by non-collaboration members.

\subsubsection{Preparation lessons learned}
\label{sec:atlas_lessons_learned}
In preparing these samples, a variety of problems were identified that resulted in improvements to the ATLAS internal metadata and software. There were also new developments specifically in support of the Open Event Generation data which will benefit ATLAS internally. For example:

\begin{itemize}
    \item inconsistencies in the number of events available in Rucio and logged in the ATLAS metadata catalogue were identified and corrected; 
    \item an issue with units in some generators that resulted in inconsistent output (e.g.~files with apparent 6.5~PeV beams, instead of 6.5~TeV beams) was identified and corrected;
    \item a simple application to translate the internal ATLAS EVNT format into \hepmc format was developed, alongside improvements to ensure that the metadata for these samples are correctly recorded. This provides a helpful option for ATLAS members wishing to debug event generator output, as the EVNT format is not trivially readable without ATLAS software;
    \item a rare bug that results in an incorrect beam energy from the event generator was identified; 
    \item numerous improvements to event generator metadata have been made.
\end{itemize}

These ancillary benefits to the collaboration help underscore the value of Open Data. In addition, as documentation of samples is made public and re-written at a level non-collaboration members can understand, the same documentation can serve as a starting point for new and non-expert collaboration members. 

\subsubsection{Future releases}

These samples represent a starting point that the collaboration has agreed upon, which will eventually total close to 14~billion events, but without any guarantee of future releases of samples or updates. With well-motivated requests, it might be possible to release additional events, additional samples, or additional data formats. For example, matrix elements in Les Houches Event (\lhe) format could be released in a similar manner, see Section~\ref{sec:partonlevel}. The total release to date is about \SI{10}{\percent} of the internally available event generator data. 

Disk resources are the primary technical constraint for the current release. Consequently, it is somewhat unclear, so far, what future releases might entail (e.g.~if necessary to save disk resources, older datasets could be replaced, even though Open Data is generally thought of as permanently available after its release).

%%%%%%%%%%%%%%%%%%%%%%%%%%%%%%%%%%%%%%%%%%%%%%%%%%%%%%%%%%%%%%
\subsection {Event generation information in CMS Open Data}
%%%%%%%%%%%%%%%%%%%%%%%%%%%%%%%%%%%%%%%%%%%%%%%%%%%%%%%%%%%%%%

The CMS experiment Open Data includes simulated data associated with, and in the same formats as, collision data released by CMS. These releases comprise all of LHC Run 1 proton-proton data at $\sqrt{s}=7$ and $8$~TeV, half of Run 2 data from 2015 and 2016 at $\sqrt{s}=13$~TeV~\cite{CMS_Open_Data_Pages}, and in addition heavy-ion (proton-lead and lead-lead) data at various $\sqrt{s}$. These simulated data are essential for research-level studies on CMS Open Data. They are processed through several steps, i.e.\ event generation, full detector simulation, signal and trigger modelling and the addition of ``pile-up'' events (simultaneous collisions at a single beam crossing) in order to make them correspond to the detector output from collisions. As the final data processing step, they are reconstructed with the same software as the raw collision data into a format convenient for physics analysis. In the context of this report, aspects related to event generation information are discussed. 

\subsubsection{Data formats}

At the time of writing, CMS has provided more than 50,000 simulated samples in various formats, including large samples of SM processes and a wide variety of BSM processes. These formats match those used in CMS analyses and vary in complexity, size, and processing requirements (e.g., needing CMS-specific software and tools). The physics content of the samples reflects the diversity of the CMS physics program and the analyses conducted with the data. The samples therefore include signal and background samples at all centre-of-mass energies and using various generators. 
Starting with the release of the 2016 collision data, data have also been provided in \nanoaod and \nanoaodsim formats~\cite{Rizzi:2019rsi} with the advantage, along with smaller event size, of not requiring CMS-specific software tools for access and analysis. As of this writing, there are 21,575 such simulation samples available with various physics processes, event generators, and sizes. The generated event information in these samples is a resource for studies that do not require CMS detector conditions. Sample sizes range up to tens of millions of events for large SM background samples, the largest exceeding 200 million events.

The \nanoaodsim data format stores event generator information for many simulated particles. It aims to provide a full description of the primary collision products by storing all particles produced by the original generator, as well as storing all charged or neutral leptons; W, Z and H bosons; tau lepton decay products; and b and c hadrons. However, the intermediate steps of the hadronisation chain are not stored, and the numerical precision of decimal quantities is limited. Generated-particle jets corresponding to the reconstructed small-radius and large-radius jets are stored, as well as generator-level MET and vertices. 

The resulting information in \nanoaodsim is much smaller than full generator information but it includes variables that are relevant to most experimental studies.
The typical \nanoaodsim event size is a few kilobytes, including both generated and reconstructed quantities. 
Where adequate, CMS simulated \nanoaodsim samples thus offer a compact, convenient data format. More complete generator information is available in larger CMS data formats such as \miniaod (20x larger than \nanoaod), which require dedicated software for analysis.  

\subsubsection{Data curation and availability}

CMS datasets are findable within the CERN Open Data Portal via text search, and using default search categories which include experiment, data format, collision type (e.g.\ proton-proton), collision energy, and category (e.g.\ ``Higgs Physics'', ``Standard Model Physics''). Each dataset has its own DOI so that its re-use can be tracked for impact on the scientific community. For each simulated dataset the full dataset name (as it appears on the CERN Open Data Portal) encodes information on the dataset itself, including the production and decay process, ranges of parameters, tune, centre-of-mass energy, and generator. A guide to decoding this information is available on the CERN Open Data Portal~\cite{CMSSimNames}.

\subsubsection{Metadata}

Each CMS Open Data record on the CERN Open Data Portal is accompanied by metadata, including:

\begin{itemize}
    \item physics keywords;
    \item a full production history with generator parameters; production scripts including generator card files, software used, and their versions;
    \item cross section values (as calculated by the event generator) with uncertainties, and filter efficiencies (if filters have been used);
    \item the centre-of-mass energy;
    \item the number of events;
    \item the variable list (i.e.\ a list of variables in the dataset given by object name, type, and a description, e.g.\  \texttt{"Electron\_pt", "Float\_t", "transverse momentum"});
    \item and a reference to software containers recommended for analysis.
\end{itemize}

While the overall discoverability of CMS Open Data is ensured by serving them through the Open Data Portal, identifying the samples of interest can be challenging. The current search facilities rely on existing metadata entries, and only provide fairly high-level categorisation. The full event generator input parameters are available, and can be downloaded for each sample, but they are not indexed for search.

\subsubsection{Usage}

The \nanoaodsim data format is stored in plain ROOT files and the generator information is readily available in a set of index-matched arrays that store quantities such as particle ID or momentum for every generated particle that was stored. Each \nanoaodsim dataset record features a variable list defining the quantities that can be accessed for each event~\cite{NanoAODSIMexample}. 

Arrays beginning with string \texttt{Generator} or \texttt{LHE} store information from the earliest stage of simulation, e.g.\ the initial colliding partons, or particles produced in the Les Houches Event format by the event generation program. Arrays beginning with \texttt{GenPart} store information for more generated particles, including selections of the particles that are produced by secondary simulation steps such as \pythia~\cite{Bierlich:2026syi}. This ``collection'' of arrays are matched by index, such that the first element in each array refers to the same particle. A ``mother'' index array is included so that decay chains can be studied, e.g.~a particle at index 6 of the \texttt{GenPart} arrays may list its mother as the particle at index 3. Each particle has one unique mother, and multiple ``daughter'' particles can point back to the same mother particle. Other arrays in \nanoaodsim files that begin with \texttt{Gen} provide information about generator-level physics objects. Detailed examples of interacting with \nanoaodsim files were presented in the 2024 CMS Open Data Workshop~\cite{Workshop2024}. 

While \nanoaodsim provides a lightweight interface to generator information, the files are not currently supported as input for fast simulation or phenomenological analysis tools. As a lightweight format, samples cannot be used for testing different jet algorithms, for example, or used for jet grooming studies. The more complex \miniaodsim format released alongside \nanoaodsim is supported by the fast simulation program \delphes. The required CMS software framework is provided via a Docker container, and \delphes can be compiled within this environment. The \texttt{DelphesCMSFWLite} reader is then available to process \miniaodsim files as a source of generated particles, avoiding the need to supply separate \lhe event records or to run \pythia along with \delphes. This method of repurposing the generated event information in \miniaodsim was used to create several billion events of \delphes output using the anticipated geometry of the CMS detector in the upcoming high-luminosity LHC (HL-LHC) era. Many HL-LHC sensitivity studies published for the 2019 CERN Yellow Report on HL-LHC physics and the Snowmass 2021 Community Study were based on these \delphes simulations~\cite{Dainese:2703572,CMS-PAS-FTR-22-001}. In future, a conversion tool could  be developed to extract the generated particle information in \nanoaodsim into a format that is readable by \delphes, or frameworks such as \checkmate~\cite{Drees:2013wra,Dercks:2016npn}, \madanalysis~\cite{Conte:2012fm,Conte:2018vmg}, or \rivet~\cite{Bierlich:2024vqo,Buckley:2010ar}.

\subsubsection{Future releases}

CMS releases data regularly and simulated samples including event generator information are part of the standard release products. The release of data collected in 2017 (19~fb$^{-1}$ of proton-proton collisions at 13 TeV) and the corresponding simulations is in preparation at the time of writing and will be made available via the CERN Open Data Portal in 2026. Limited releases of generator information in dedicated formats could be considered in future if the impact on the broader community would be significant, though the mission of the CMS Open Data program is focused on releasing data in formats used within CMS. 

%%%%%%%%%%%%%%%%%%%%%%%%%%%%%%%%%%%%%%%%%%%%%%%%%%%%%%%%%%%%%%%%%%%%%%%%%%%%%%%%
\section{Examples of Use Cases and User Experiences}
\label{sec:usecases}
%%%%%%%%%%%%%%%%%%%%%%%%%%%%%%%%%%%%%%%%%%%%%%%%%%%%%%%%%%%%%%%%%%%%%%%%%%%%%%%%

This section presents a selection of use cases and user experiences.
Although not exhaustive, these examples serve to illustrate the benefits of shared MC samples, to
highlight the broader potential of existing resources,
and to demonstrate approaches to improving their usability.

%%%%%%%%%%%%%%%%%%%%%%%%%%%%%%%%%%%%%%%%%%%%%%%%%%%%
\subsection{Sharing samples between experiments}
\label{sec:sharing}
%%%%%%%%%%%%%%%%%%%%%%%%%%%%%%%%%%%%%%%%%%%%%%%%%%%%

One obvious use-case for open event generation data is the sharing of samples between experiments, which is also explored by the recently established ``Data Sharing and New Workflows'' task force of the LHC MC WG~\cite{MCWG:datasharing}.
This would require agreement between experiments on a mutually compatible format, but of course, does not mandate the use of the same nominal samples across experiments. In some cases, experiments have well-motivated reasons for selecting different event generator configurations for their primary physics estimates. However, the sharing of samples allows the possibility that ATLAS can include the exact CMS configuration in a figure, that LHCb can include the exact ATLAS configuration, and so on. The resources devoted to event generation by the experiments are substantial, and real and extensive sharing could provide significant financial and environmental benefits. These benefits are explored further in Section~\ref{sec:sustainability}.  

A successful precedent exists: in Ref.~\cite{ATL-PHYS-PUB-2023-016} ATLAS and CMS agreed on a configuration for top-quark production samples with considerable effort, and subsequently ATLAS generated \lhe files with \powheg~\cite{Frixione:2007vw,Alioli:2010xd} that were shared via a CERN file storage system. Both experiments then showered the events using the same programs and settings in order to create identical samples in their respective production systems. The experiments also agreed on settings for \sherpa samples and independently generated samples with those settings. Open event generation data could provide a significantly more straightforward means by which the samples could be trivially shared, including with the broader community that might not have access to the CERN file storage system. There are also cases where the generation itself is rather complex (e.g.~dark showers samples), and sharing samples ensures that all groups are examining the same physics.

Using the same nominal samples also raises some concerns, such as the possible introduction of statistical and systematic correlations between results from the two experiments, or the reproduction of errors. Independent generation with consistent setups, or assigning each experiment half of a common sample to use, would statistically de-correlate the samples, but would reduce the number of events available to each experiment, and the samples would still have identical systematic uncertainties.

%%%%%%%%%%%%%%%%%%%%%%%%%%%%%%%%%%%%%%%%%%%%%%%%%%%%
\subsection{Reuse and reinterpretation of measurements}
\label{sec:ReuseMeasurements}
%%%%%%%%%%%%%%%%%%%%%%%%%%%%%%%%%%%%%%%%%%%%%%%%%%%%

Tools such as \rivet~\cite{Bierlich:2024vqo,Buckley:2010ar} and \contur~\cite{CONTUR:2021qmv,Butterworth:2016sqg}
exploit particle-level measurements of differential cross sections to validate and tune new SM predictions and
to test BSM scenarios. The MC event samples released by experiments allow additional flexibility and robustness in such studies, 
for example by allowing the baseline SM predictions used by one experiment to be compared to measurements from the other. These
comparisons can then in principle be used to investigate any discrepancies, to probe BSM models, or to perform fits of SM or effective field theory
parameters.

\begin{figure}[t]\centering
    \includegraphics[width=0.48\textwidth]{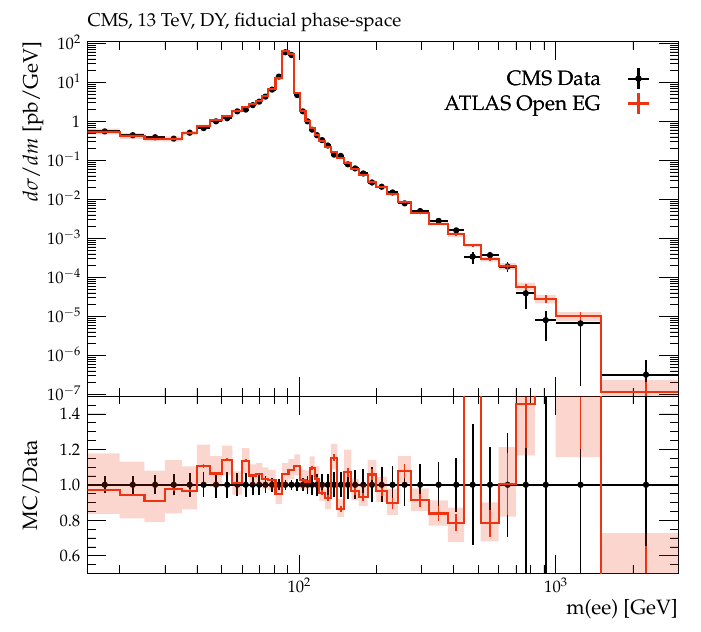}\includegraphics[width=0.48\textwidth]{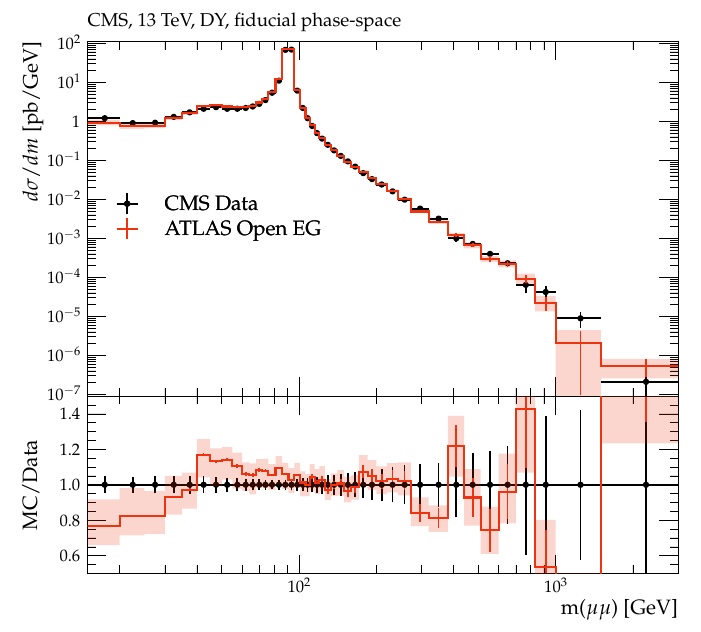}
    \caption{ (a) CMS Drell-Yan di-electron and (b) dimuon cross section
      measurements compared to ATLAS Open Event Generation (\sherpa).}
      \label{fig:CMS-DY-ATLAS-OD}
\end{figure}

A proof-of-principle example is shown in Figure~\ref{fig:CMS-DY-ATLAS-OD}, where the CMS
high-mass Drell-Yan measurement~\cite{CMS:2018mdl} is compared to a prediction made using
the recently-released ATLAS dilepton samples produced with \sherpa.
A fully automated pipeline~\cite{hepmcaccess} was developed in the process of this study, using \rivet and HTCondor~\cite{HTCondorDocs}.
The pipeline filters the selected \sherpa Drell-Yan datasets ($Z\to ee$, $Z\to\mu\mu$, and $Z\to\tau\tau$, across $p_T^V$ slices) using the metadata with a
dedicated script, and analyses each dataset independently with \rivet on a batch system, while computing the cross section normalisation factors,
calculated simultaneously as 
\begin{equation}
\mathrm{normalisation} = \sigma \times k\text{-factor} \times \epsilon_{\mathrm{filter}} \, .
\end{equation}
It then merges each different \yoda~\cite{YODA} output file, using the metadata associated with the relevant \hepmc samples.
Since \rivet routines internally perform the normalisation necessary to produce a differential cross section, one does not need to
divide by the sum of weights or bin width. However, in calculating the uncertainties, the appropriate weight variations (in this case, different
parton density and scale variations) are required.

The \texttt{atlasopenmagic} metadata API~\cite{atom25} was found to be user-friendly and flexible, enabling efficient filtering of the relevant datasets.
Some issues encountered along the way are discussed below.

In the Open Event Generation samples used for the calculations, the ATLAS metadata gives a ``None'' output for the weights, however the (hundreds of) weight variations
are present in the \hepmc files.
They are labelled by numbers, and so an external look-up table was requested (and provided) to map these to the \sherpa labels,
which follow the standard convention~\cite{Bothmann:2022weight}.
This allowed the weights to be interpreted correctly; in particular the three unphysical weights \sherpa uses (UserHook, NTrials, and WeightNormalisation)
and also the internal matrix-element weight, MEWeight,
had to be excluded. Since \texttt{atlasopenmagic} 1.9.0, users can fetch the weight information in a straightforward manner~\cite{atom_weights}.
Looking ahead, it would be beneficial if Open Event Generation weight names in the \hepmc file followed the standard names and ordering~\cite{Bothmann:2022weight}. This would simplify using weights across different event generators, reduce duplication of efforts, and minimise the risk of analysis-level errors.
Named weights are supported by the \hepmc~3 data format, though they are not written by default.

The unit issue described in Section~\ref{sec:atlas_lessons_learned} was also identified during this work.
Regarding the rare bug resulting in incorrect beam energy described in Section~\ref{sec:atlas_lessons_learned} and identified in this analysis,
an option to flag such occurrences and continue (rather than terminating the run) will be implemented in  \rivet.
When fetching a large number of datasets (of the order of ~5000), using the released version resulted in the cache filling up very quickly.
A newer (at the time beta) version of \texttt{atlasopenmagic} (v1.7.0) overcame this problem.

Finally, the \hepmc outputs state the library version as  \texttt{HepMC::Version 3.03.00}.
However, the file format used is \texttt{HepMC::IO\_GenEvent} which corresponds to the legacy \hepmc~2 format.
The present framework has been adapted to support the \hepmc~2 format, but it would be preferable to move to using \hepmc~3 format in future as this is better supported.\footnote{See also Section~\ref{sec:atlas_dataformat} for comments on the \hepmc~2 and \hepmc~3 formats.}

While some challenges were encountered in understanding the metadata and accessing the events, the outcome was successful, and several improvements to enhance clarity, user-friendliness and consistency across different measurements were made, or are in progress.
This demonstrates a proof-of-principle for the Open Event Generation approach, and the predictions shown in Figure~\ref{fig:CMS-DY-ATLAS-OD} will be added to the \hepdata record associated with this paper, and deployed, for example, in future releases of \contur.

%%%%%%%%%%%%%%%%%%%%%%%%%%%%%%%%%%%%%%%%%%%%%%%%%%%%
\subsection{Studying overlaps across reinterpreted signal regions}
%%%%%%%%%%%%%%%%%%%%%%%%%%%%%%%%%%%%%%%%%%%%%%%%%%%%

To fully exploit the LHC dataset, it is interesting to combine signal regions across different analyses within the same experiment.
Accordingly, reinterpretation efforts such as those described in Section~\ref{sec:ReuseMeasurements} rely on a consistent statistical treatment across measurements, where neglected correlations can bias likelihood estimates.
This requires a reliable understanding of inter-analysis correlations, which is non-trivial due to the complex and highly structured selection criteria employed in modern analyses.

The \rivet\!+\contur pipeline mentioned in Section~\ref{sec:ReuseMeasurements} uses a heuristic grouping strategy in which similar final states are pooled, and it is  conservatively assumed that histograms within a pool are not independent.
An alternative automated approach to the problem is to estimate overlaps directly by evaluating the co-occurrence of events across signal regions using large samples of simulated events, as suggested in~\cite{Araz:2022vtr}.
This enables a data-driven determination of correlations.
Such an approach requires event samples that are both sufficiently large and representative.
Standard Model samples provide an unbiased baseline for estimating typical overlaps~\cite{Yellen:2025ryl}.
Conversely, a diverse set of BSM samples, particularly those targeting extreme or low-background regions, allows correlation structures relevant for reinterpretation to be probed, similarly to the set of SUSY searches studied in Reference~\cite{Feike:2024zfz}.
The required scale and diversity make Open Event Generation datasets uniquely suited to this task.

\begin{figure}[!t] \centering
        \includegraphics[width=\textwidth]{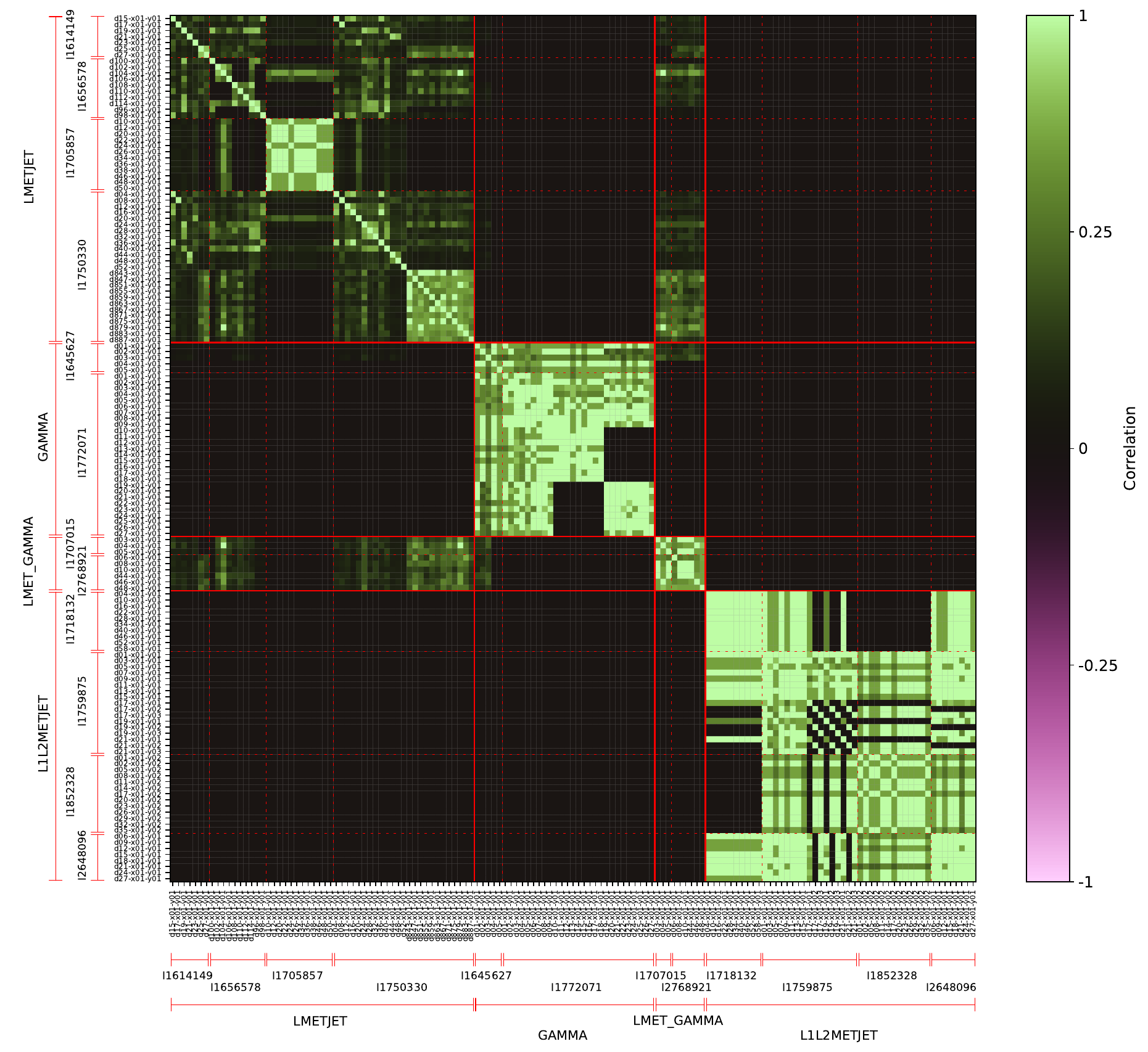}
    \caption{
    Correlations across all non-empty histograms from 13~TeV ATLAS measurements used by default by \contur{} (excluding ratio measurements) calculated on a BSM sample from the ATLAS Open Event Generation release (sample \#310780).
    \contur{} pools are delineated by solid red lines: as expected, the majority of the significant correlations lie within the ``same-pool'' block diagonal. Correlation across pools indicates that the combination may have been too aggressive, whereas non-correlation within the diagonal blocks offers the possibility of setting stronger limits by more aggressive combination.
    The twelve individual analyses \cite{ATLAS:2017cez,ATLAS:2018acq, ATLAS:2018fwl, ATLAS:2019hxz, ATLAS:2017xqp, ATLAS:2019iaa, ATLAS:2018sos, ATLAS:2024hmk, ATLAS:2019ebv, ATLAS:2019hau, ATLAS:2021jgw, ATLAS:2023gsl} are labelled by their Inspire IDs and delineated by thinner dashed lines: significant correlation between different analyses within the same pool is to be expected.
    }
  \label{fig:Rivet_Correlations} 
 \end{figure}
 
Preliminary results using this approach are shown in Figure~\ref{fig:Rivet_Correlations}, which presents the correlations between \rivet{} histograms for a BSM model available in the ATLAS Open Event Generation release (\#310780).
The analyses are grouped according to their \contur{} pools to allow comparisons with the heuristic grouping.
The largely block-diagonal nature of the correlation matrix indicates that most significant correlations occur within pools, demonstrating that the \contur{} approach captures the dominant correlation pattern.
At the same time, the presence of near-zero correlations within several pools suggests that more aggressive combinations, treating some within-pool histogram pairs as independent, could yield stronger exclusion limits without introducing significant bias.
Conversely, the small but non-negligible cross-pool correlations in Figure~\ref{fig:Rivet_Correlations}, while not affecting the final likelihood in this particular example, indicate that the current pooling strategy is not uniformly conservative.
Open Event Generation enables a transition from assumption-driven treatment, that can  occasionally be optimistic, to data-driven handling of signal region overlaps, with direct implications for both search design and reinterpretation.

The use of Open Event Generation data in this study was relatively smooth.
The BSM sample used was straightforwardly accessible via \texttt{atlasopenmagic}, and the \rivet{}+\contur pipeline described in Section~\ref{sec:ReuseMeasurements} required only minor adaptation.
Extending this study to cover the full set of ATLAS BSM Open Event Generation samples, and ultimately to SM processes, would allow a statistically robust and increasingly model-independent map of inter-analysis correlations across the ATLAS measurement programme.
These extensions are currently limited by the need to download large datasets locally, since the tarred \hepmc files cannot yet be streamed directly via \xrootd.
Including CMS Open Event Generation would further broaden coverage and improve robustness.
However, interfacing the CMS \miniaod and \nanoaod formats with reinterpretation tools is currently not possible; provision of CMS Open Event Generation in the community-standard \hepmc\ format would therefore be advantageous.
Resolving these two technical bottlenecks would be the most impactful near-term improvements for this use case.

%%%%%%%%%%%%%%%%%%%%%%%%%%%%%%%%%%%%%%%%%%%%%%%%%%%%
\subsection{Designing new BSM searches}
\label{sec:DesignNewBSMSearches}
%%%%%%%%%%%%%%%%%%%%%%%%%%%%%%%%%%%%%%%%%%%%%%%%%%%%

Collider searches often require the simulation of SM backgrounds, which is carried out independently by each experimental collaboration, or on a publication-by-publication basis by non-collaboration authors. Backgrounds generated can have large cross sections, requiring significant computational power to generate sufficient statistics, and significant expertise to accomplish correctly, due to non-trivial generator settings or subtleties of merging/matching, often requiring multiple attempts at generation. These issues will be exacerbated in higher precision studies and with increasing luminosity, placing an increasing burden on a small group of MC developers to troubleshoot the usage of a large community of non-experts. 
Large shared event samples, generated by experts, are an obvious boon in this context. 

An example analysis 
benefitting from ATLAS Open Event Generation data is a proposed % BSM
search in the opposite-sign dimuon + jets + missing energy channel, with 10~GeV $\le M_{\mu^+\mu^-} \le 50$ GeV \cite{Mahbubani:inprep}, at 13.6 TeV centre-of-mass energy.  The search was designed to probe a two-component scalar dark matter model with one active and two inert Higgs doublets, and a $Z_2\times Z'_2$ symmetry, at an integrated luminosity of 300~fb$^{-1}$.  Beside the usual SM backgrounds, $(W\to\mu\nu)+c$ was expected to provide a significant source of fake/non-prompt muons, with the non-prompt muon from the charm decay mistagged as a second prompt, isolated muon.  The cross section for $(W\to\mu\nu)+c$ at 13.6 TeV is large: 3.4~nb at NLO; this corresponds to over one billion events for 300~fb$^{-1}$.  Compounding the challenge of simulating large statistics is a need for accurate modeling of the jet activity and spectrum, and the missing energy, to which the fake rate estimate will be sensitive.

The ATLAS $(W\to\mu\nu)+c$ ``Baseline'' dataset at 13.6~TeV centre-of-mass (\#700781) consists of 50 million uploaded events (of a total of 373 million generated by ATLAS); insufficient statistics even if the full dataset were made available.  Instead, ``specialised'' dataset \#501719 at $\sqrt{s}=13$~TeV, consisting of $W\to\mu\nu+0,\,1,\,2,\,3$~jets at NLO, with a multi-muon filter that captures heavy flavour decays to muons, generated with \madgraph~\cite{Alwall:2011uj,Alwall:2014hca}, and showered with \pythia, was used instead. 
This sample had a smaller cross section, 76.9~pb, with sufficient statistics to reach an event weight of just over 2 for the fake rate estimate.\footnote{Although many of the available datasets, like the $W$ sample used here, contain limited statistics, it is possible to request more events if a strong case can be made for them.  It is also sometimes possible to combine a ``baseline'' dataset with different ``specialised'' samples, produced for systematics estimation using different event generators or parton showers. However this option was not available for $W+X\to\mu\mu$, itself a ``specialised'' set.
}  
The result was then rescaled to $\sqrt{s}=13.6$~TeV. 

The dataset was large, consisting of a total of 10M events, distributed across 1000 tarred and gzipped \hepmc files of just over 500 MB in size, each taking up to 2 mins to download (about 33 hours total). A custom wrapper, based on the reference tutorial, was written to facilitate bulk data download.  The socket timeout threshold was increased to 120~s to allow for network latency, with intermediate status reporting to track progress. The progress information was indispensable, particularly in light of the slow download speeds, and frequent download errors due to server timeout, and we manually verified which files in the dataset were not downloaded successfully in order to attempt to re-download them.  

\begin{figure}[t]
\centering
\includegraphics[width=0.66\textwidth]{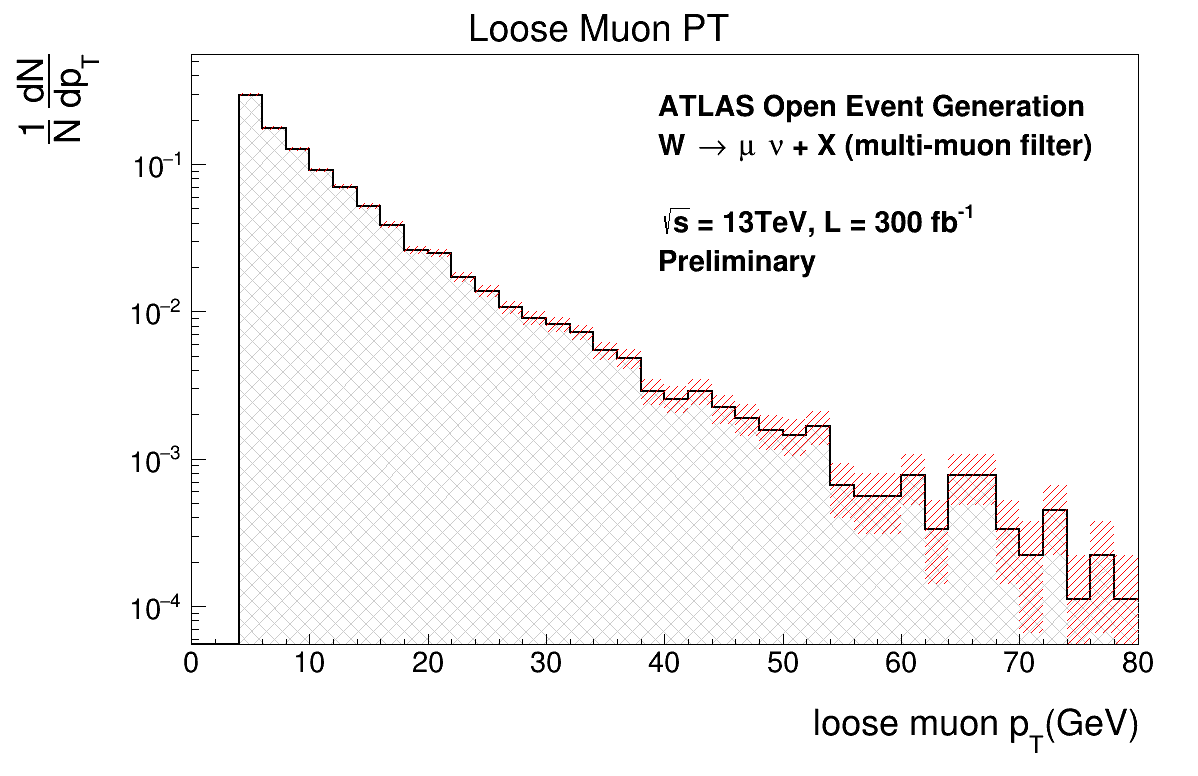}
\caption{Preliminary plot of ``loose'' muon $p_T$ in a control region consisting of two same-sign muons satisfying all preselection cuts for a proposed LHC search for a two-component scalar dark matter model, for $\sqrt{s}=13$~TeV and an integrated luminosity of 300 fb$^{-1}$. This control region will be rescaled to $\sqrt{s}=13.6$~TeV and used to estimate the rate of non-prompt muon backgrounds to the search.\label{fig:CR-SSmuons}}
\end{figure}

The \checkmate~\cite{Dercks:2016npn} framework interfaced to \delphes~\cite{deFavereau:2013fsa} was used in the analysis design, allowing for convenient access to customised object selection, e.g.~baseline muons. The fake/non-prompt background rate was estimated from a simplified fake factor method~\cite{Aad_2023}, derived using a ratio of ``loose'' (baseline muons with no isolation or overlap requirements) to ``tight'' muons in a control region. 
Figure~\ref{fig:CR-SSmuons} shows a preliminary plot of ``loose'' muon $p_T$ in a control region with two same-sign muons satisfying standard preselection cuts. 

Overall, the \texttt{atlasopenmagic} tool was easy and intuitive to use, and all queries received swift responses from the ATLAS Open Event Generation team via the CERN Open Data Forum~\cite{od-forum}. Nonetheless it can be relevant to discuss two issues encountered. 
First, 
\checkmate produces cutflow tables for each individual \hepmc file, which need to be combined manually for a given process.  For this particular dataset, occasional negative weights from the NLO generation, in tandem with the small-statistics \hepmc files (10,000 events per file is the ATLAS standard for Open Event Generation), resulted in several unphysical cutflow outputs, with negative or increasing event numbers with subsequent cuts.  Combining the cutflow results from the full dataset, however, provided physically sensible output.  

Second, running \checkmate on the files required untarring and unzipping each file: with an unzipped file size of ~2 GB, this required 2 TB of hard disk for storage of a single dataset (with a factor of 2 too small statistics for this particular analysis), making storage space a limiting factor. 

The large size seems partially due to multiparton interactions (MPI) being turned on by ATLAS during showering and hadronisation, and partially from the use of specific MC event generators, which write to \hepmc differing amounts of information.\footnote{As a benchmark, a \hepmc file containing semileptonic $t\bar{t}$ events generated by ATLAS using \sherpa plus \pythia with MPI turned on was approximately 65\% larger than the same process when generated using \madgraph plus \pythia with MPI turned off.}  Default \hepmc file sizes could be substantially reduced with minimal information loss by decreasing the default precision to which particle 4-momenta are written, and decreasing the amount of intermediate information written to file.  Discussions of how to do the latter systematically across the different event generators are currently under way.  Regardless,  inclusion in the metadata of the total number of files in each Open Event Generation dataset, and its compressed and uncompressed size, would make it easier to ensure sufficient storage for its download and use.  

In the long term, allowing users access to the data and all necessary downstream tools via a central analysis facility (see Section~\ref{sec:CAF}) with a `light' user interface via e.g.\ a python notebook and output histograms, would bypass many of the problems associated with local download and storage required when streaming the files via \xrootd is not supported. 
Near-data analysis would be less resource- and carbon-intensive than streaming via \xrootd and would allow easier access for users without a reliable internet connection.

Quantification of fake rates, like other effects that rely on accurate modelling of particle-detector interactions (e.g.~punch-through of jets from the calorimeter into the muon system and its effects on jet resolution) can be difficult to reproduce by piping Open Event Generation data through parametric detector simulations like \delphes, which are generally tuned using public information. 
Similarly, long-lived particle (LLP) searches that rely on the LLP interactions with the detector, and often involve low-level detector variables, are difficult to emulate using Open Event Generation data alone.
An additional complexity comes with the use of machine learning in LHC analyses, which can introduce strong dependence on aspects of the data that are not described by parametric simulation.\footnote{In some cases, for example where the signal is selected with a reasonably high efficiency that is relatively easy to reproduce based on particle-level information~\cite{SUSY-2016-16}, or where a surrogate model is provided by the collaboration specifically to overcome this difficulty~\cite{MLinAnalysis1,MLinAnalysis2}, reinterpretation is still possible with Open Event Generation data alone.}  Although not strictly a limitation of the Open Event Generation data itself, this motivates the enhancement of the Open Data set by the inclusion by experiments of open ``event generation'' data with detailed detector modelling effects based on the output of their internal detector simulations.  This could include datasets with BSM states with non-prompt/non-standard decays that are not part of event generation output, such as $R$-hadrons, or other LLPs.

%%%%%%%%%%%%%%%%%%%%%%%%%%%%%%%%%%%%%%%%%%%%%%%%%%%%
\subsection{Proof-of-concept for a novel quark/gluon-jet measurement}
%%%%%%%%%%%%%%%%%%%%%%%%%%%%%%%%%%%%%%%%%%%%%%%%%%%%
A proof-of-concept study of the novel quark/gluon-jet measurement proposed in Ref.~\cite{Baron:2023hkp} was performed using CMS Open Data. 
The method relies on measuring the same jet-substructure observable in inclusive jet samples at two (or more) different centre-of-mass energies and was validated in \cite{Baron:2023hkp} using MC event samples generated with \herwig~\cite{Bahr:2008pv,Bellm:2015jjp,Bewick:2023tfi,Bellm:2025pcw} and \pythia. CMS Open Data make it possible to apply the strategy to real collisions thanks to the availability of detector-level reconstructed information in the form of Particle-Flow (PF) candidates, as well as generator-level information that enables unfolding. 

The data used in this study are taken from the CERN Open Data Portal and correspond to jet primary datasets collected at $\sqrt{s}=7$~TeV, $8$~TeV, and $13$~TeV~\cite{CMS_Dataset_7TeV,CMS_Dataset_8TeV,CMS_Dataset_13TeV}.
Although jets at all energies are clustered with the anti-\(k_t\) algorithm~\cite{Cacciari:2008gp} as implemented in \textsc{FastJet}~\cite{Cacciari:2011ma}, the main complication is the change in the jet definition at \(13\)~TeV relative to lower energies: the jet radius parameter is reduced from \(R=0.5\) to \(R=0.4\), and the Charged Hadron Subtraction (CHS) procedure~\cite{CMS_PileupMitigation} is introduced.
To harmonise the jet definitions across centre-of-mass energies, the authors reanalysed the lower-energy datasets using the \(13\)~TeV jet definition, i.e.\ anti-\(k_t\) jets with \(R=0.4\) clustered from all charged and neutral PF candidates, with pileup contributions mitigated via CHS.

The analysis also includes an unfolding procedure to correct reconstructed distributions for detector effects and thereby infer the corresponding particle-level observables~\cite{Cowan:2002in,Hocker:1995kb,Adye:2011gm}. This step is essential for producing measurements that are less dependent on a specific detector implementation and can therefore be compared directly with theoretical predictions and with results from other experiments~\cite{Cowan:2002in,Adye:2011gm}. 
Although the measurement is based on collision data, MC event samples remain indispensable, since they are used to construct the response matrix and to determine acceptance and efficiency corrections linking reconstructed and true observables \cite{Hocker:1995kb,Cowan:2002in,Adye:2011gm}. Several CMS multijet simulations such as Ref.~\cite{CMS:2024YYED6QT9} were used to perform the unfolding, with ``generated jets'' formed from clusters of generated particles providing the particle-level information corresponding to reconstructed jets in the simulation. Open CMS event generation data therefore played a central role in the study, which was recently presented in~\cite{Epiphany2026}. 

%%%%%%%%%%%%%%%%%%%%%%%%%%%%%%%%%%%%%%%%%%%%%%%%%%%%
\subsection{Machine learning developments}
%%%%%%%%%%%%%%%%%%%%%%%%%%%%%%%%%%%%%%%%%%%%%%%%%%%%

The development and evaluation of ML models often depend on labelled datasets, which in HEP typically means simulated data. As such, several teams have produced dedicated simulations, for instance, as challenges to promote ML developments \cite{Kasieczka:2021xcg,Krause:2024avx,Adam-Bourdarios_2015,Chakkappai:2025noy,Aarrestad:2021oeb} or more recently for large-scale training \cite{Qu:2022mxj,McKeown:2025gtw,elitez2025collidermlreleaseopendatadetectorhighluminosity,bhimji2025fair,Amram:2024fjg}. However, as both the scale of data required and the variety of potential applications grow rapidly, the direct use of event simulation produced by LHC collaborations becomes increasingly valuable. To illustrate this, we describe four use cases in which open LHC MC event simulation has been or would be of great use: benchmarking, development of foundation models, scaling studies, and mitigation of negative event weights.

\paragraph{Benchmark datasets:} Dedicated simulation datasets tailored to showcase machine learning applications are essential for efficient and reproducible ML development in HEP. Typically, an LHC collaboration produces a specific simulation and uses it to develop and test a given ML method; however, as new techniques emerge, it is far more efficient and transparent for the community to reuse the same datasets rather than reimplementing and rerunning the full MC chain, which can take months. This need is exemplified by cases such as the recent CMS di-jet anomaly detection analysis \cite{CMS:2025sch}, where recently released datasets now allow external researchers to benchmark new methods without repeating the simulation. Prior to this release, subsequent studies were required to independently reproduce the full simulation in order to compare their results with those of the original analysis (see for instance Ref.~\cite{Das:2026spe}). Another example is FlashSim~\cite{Vaselli:2024hml}, an ML tool developed within the CMS collaboration for the direct generation of reconstructed physics object information from generator-level input. The original work relied on full GEANT4~\cite{GEANT4:2002zbu} simulation. However, subsequent developments that improved the performance of the method~\cite{Vaselli:2024vrx} and proposed new generative network validation techniques~\cite{Cappelli:2025myc} required the use of light-weight \delphes~\cite{deFavereau:2013fsa} simulations produced independently. Fully open and standardised Monte Carlo benchmark datasets would therefore reduce duplication of effort and enable more efficient, transparent, and comparable ML development across the field.  They can also be used to validate new and existing generators.

\paragraph{Foundation models:} The development of foundation models relies on pre-training on large amounts of data, followed by re-training on specific downstream tasks. 
In this context, both real data and simulation play important roles. While most scientific fields rely almost exclusively on real data for pre-training due to the lack of sufficiently realistic simulations, particle physics benefits from vast quantities of high-fidelity MC simulation. This creates a unique opportunity to study, e.g., how pre-training-plus-fine-tuning strategies scale relative to fully supervised training at a level unmatched in other domains.

Efforts have been made to produce large datasets for pre-training transformer networks on domain-specific tasks, such as jet events and calorimeter showers \cite{Qu:2022mxj,McKeown:2025gtw,elitez2025collidermlreleaseopendatadetectorhighluminosity,bhimji2025fair,Amram:2024fjg}. However, the bulk of available data comes from the LHC collaborations, both as real, observed data, and as simulations. The efficient use of event-level data often requires preprocessing pipelines to make them readily usable for training ML methods on specific tasks. For example, the Aspen Open Jets (AOJ) ML-ready dataset was extracted from CMS Open Data, comprising 180 million jet events and converted from the \miniaod format to an \hdffive tabular format, with the goal of enabling foundation model development~\cite{Amram:2024fjg}. This dataset has since been complemented with $\sim$300 million CMS simulation events of QCD and top jets~\cite{MCAOJ:2026}. Notably, even at the pre-training stage, including simulation alongside real data may be relevant in collider physics. 

Relying only on real data may introduce biases, whereas pre-training on diverse simulated data can allow models to learn broader representations, including potential new physics signatures. This remains an active area of study. An example is given in Ref.~\cite{Bhimji:2025isp}, where a diverse dataset of order one billion jet events, combining real LHC data \cite{Amram:2024fjg}, collaboration-independent simulation \cite{Li:2024htp,Li:2024htp} together with collaboration-produced QCD and BSM simulations from the CMS collaboration, neutral-current deep inelastic scattering (DIS) simulations from H1 and the ATLAS top-tagging dataset \cite{ATLAS:2024rua}, was used to train a foundation model that achieved a 30\% improvement over state-of-the-art performance in top tagging.

\paragraph{Scaling laws:} The limits of ML methods as the available data scales is a field of growing interest globally, and has permeated into HEP. In a recent study performed within the ATLAS collaboration \cite{3113512}, the scaling laws of jet flavour tagging were thoroughly investigated using of the order of 10 billion ATLAS simulation events, showcasing the potential gains from scaling in data, model size, and computing time. However, since the simulation data used are not open, a very similar study was released to encourage further research \cite{Vigl:2026ppx}, using the publicly available JetClass-1 dataset \cite{Qu:2022mxj}, consisting of 100 million jet events—two orders of magnitude less data. To fully perform and reproduce scaling law studies at large scale in a transparent and reproducible manner, the release of LHC event simulation is needed. Relevant directions include scaling behaviour in anomaly detection, generative modelling, and comparisons between self-supervised and supervised foundation model pre-training strategies.

\paragraph{Event Weights:} One long-standing issue for event generators, which can be mitigated using machine learning~\cite{EvgenReweighting}, is negative event weights.  These can spoil statistical precision and make it difficult to generate sufficiently large samples.
One solution explored in Ref.~\cite{EvgenReweighting} is training neural networks to re-weight positively-weighted events to the full ensemble of events. This sort of re-weighting is trivial in a single-dimensional histogram, but attempting the same re-weighting over the full kinematic space of the events to derive a general solution to the problem is significantly more complex. Moreover, with a full kinematic space re-weighting and fine-tuning techniques, it might be possible to provide a tool that could re-weight arbitrary samples, rather than one that must be re-trained each time a new sample is generated. This is an active area of research.
These sorts of studies can be done entirely with event generator open data, and they can benefit from the diversity of samples that have been released. \\

Thus, we encourage the release and documentation of simulation data for future ML developments, including training data used for ML research within LHC collaborations. Here, not only the datasets, but also the simulation configurations (generator cards) used to produce them, as well as the trained models~\cite{Araz:2023mda,Bieringer:2024pzt}, should be made available in a systematic and accessible way.
Finally, proper converters of multi-purpose LHC event simulation from \miniaod to ML-friendly formats are needed.

%%%%%%%%%%%%%%%%%%%%%%%%%%%%%%%%%%%%%%%%%%%%%%%%%%%%
\subsection{Training and education}
%%%%%%%%%%%%%%%%%%%%%%%%%%%%%%%%%%%%%%%%%%%%%%%%%%%%

The Open Event Generation data also provide a useful basis for some training and education efforts. As a part of the documentation, for example, ATLAS has developed a simple tutorial demonstrating the use of the Open Event Generation data~\cite{ATLAStutorial}. This tutorial also explains the \hepmc data format and offers users an opportunity to look at the real output of event generators, and to understand how particle records work, what PDG identifiers are, how status codes work, what sorts of interactions are represented by vertices, and so on. Such a tutorial can be useful in disabusing students of the notion that event generator output is a ``simple'' tree of branchings, and can encourage users to think about the complexities of the internal structure of event generator records that attempt to represent quantum mechanical processes (e.g.~that it is not trivial to identify ``the top quark'' in a top-quark production event).

%%%%%%%%%%%%%%%%%%%%%%%%%%%%%%%%%%%%%%%%%%%%%%%%%%%%%%%%%%%%%%%%%%%%%%%%%%%%%%%%
\section{Financial and environmental impact}
\label{sec:sustainability}
%%%%%%%%%%%%%%%%%%%%%%%%%%%%%%%%%%%%%%%%%%%%%%%%%%%%%%%%%%%%%%%%%%%%%%%%%%%%%%%%

There is often a tension between maximizing scientific output and reducing operational and environmental costs.  
Sharing event generation data presents a counterexample: by eliminating duplicated effort and generation and storage of redundant datasets, it is possible to reduce the computational, financial, and environmental footprint of MC simulation with no adverse effect on the physics done.
This approach comes with a trade-off, however, in the form of increased network traffic, as large event files are moved or streamed over the network to the computing sites of users outside the experimental collaborations, who would otherwise have simulated the events locally; the associated infrastructure and environmental costs are not accounted for here.  The establishment of a central analysis facility enabling near-data computing, as described in Section~\ref{sec:CAF}, would go some way towards mitigating this overhead.

%%%%%%%%%%%%%%%%%%%%%%%%%%%%%%%%%%%%%%%%%%%%%%%%%%%%
\subsection{Financial impacts} \label{sec:financial-impacts}
%%%%%%%%%%%%%%%%%%%%%%%%%%%%%%%%%%%%%%%%%%%%%%%%%%%%

% \subsection{ATLAS}
The ATLAS Open Event Generation has averaged 232~HS23-seconds (HS23s) per event over the last two years~\cite{grafanaATLAS}.\footnote{The HEPScore23 (HS23) is a benchmark score designed to measure the CPU performance of computing hardware for High Energy Physics, particularly for the Worldwide LHC Computing Grid~\cite{HEPScore}. Multiplying by a unit of time gives a normalised unit of computing time, e.g.\ HS23-second (HS23s) or HS23-year (HS23y).}
The 12.8 billion generated events released therefore correspond to about 2.97 trillion HS23s, or $100$~kHS23y.
At the time of writing, the average CPU core is around 15 HS23,\footnote{This estimate corresponds to the average over the last two years of the HS23/core ratio from Refs.~\cite{hepix,wlcg_accounting}, considering simultaneous multithreading, and excluding the top and bottom \SI{5}{\percent} of the distribution.} which implies about 198 billion CPU-core-seconds (CPUs).
That is equivalent to approximately $3300$ CPU-cores, or $50~\textrm{kHS23}$, used at saturation for two years. Including a \SI{5}{\percent} failure rate in terms of wall-clock time~\cite{grafanaATLAS,Megino:2903380} implies a total of $3500$ CPU-cores, or $53~\textrm{kHS23}$.

\begin{figure}[!t] \centering
    \includegraphics[width=\textwidth]{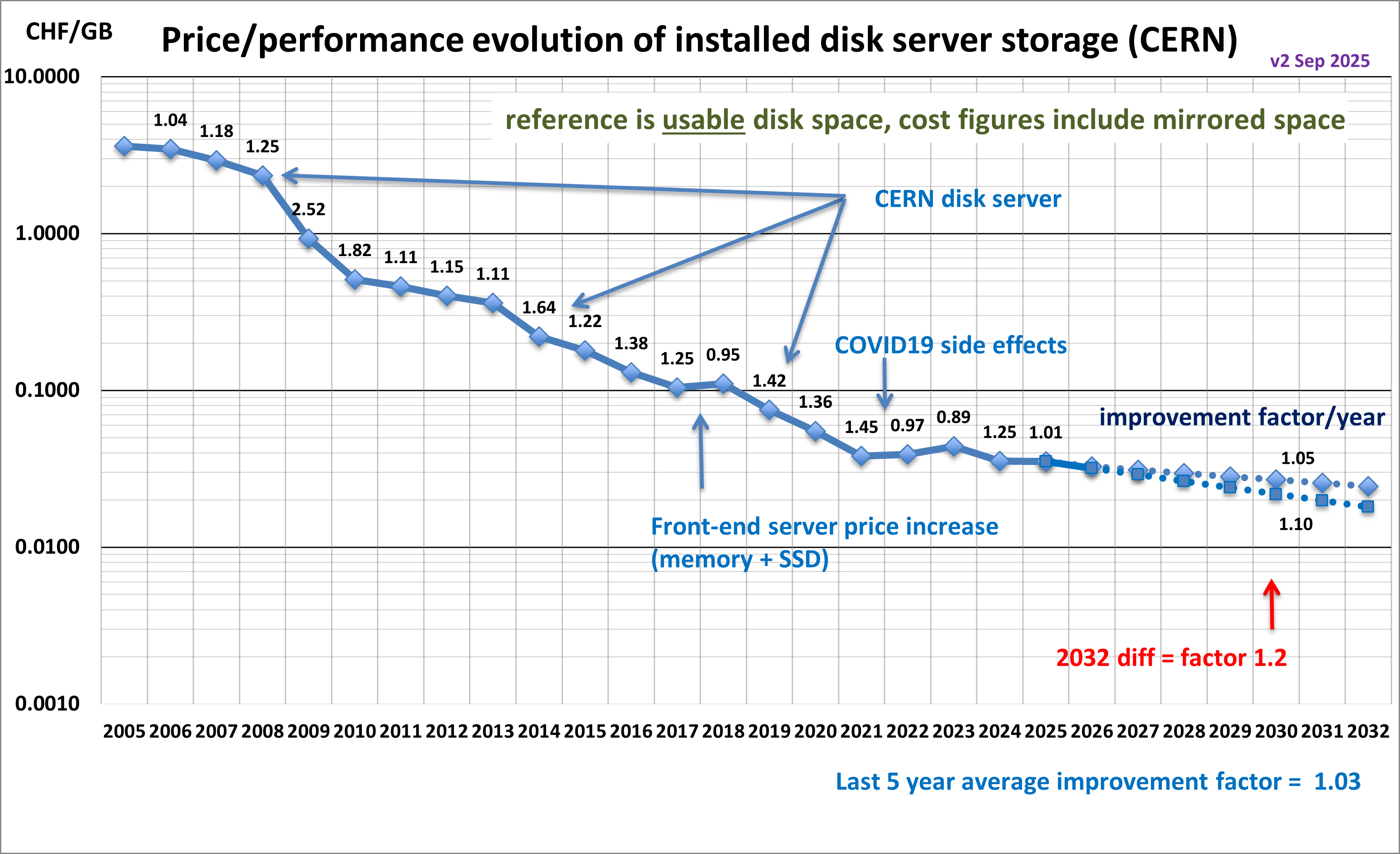}
    \caption{Evolution of price/performance for installed disk server storage at CERN (2005--2025), measured in CHF/GB usable space (including mirrored space). The projections exclude recent RAM and storage price fluctuations from AI demand. Figure taken from~\cite{sciaba}.}
    \label{fig:STORAGE-COST}
 \end{figure}

To quantify the financial impact and corresponding savings from publishing Open Event Generation samples, two hypothetical scenarios are presented below: regeneration using CERN Tier-0 infrastructure, or using Google Cloud services.
The outcome is summarised (with cost estimates expressed in thousands of CHF) in Table~\ref{tab:cost-summary}.

\subsubsection{CERN Tier 0 impact}

To account for the financial impact of compute resources at the CERN Tier 0, it is possible to estimate that $ 1\textrm{kHS23y}\simeq \text{CHF}\,1.8\textrm{k}$~\cite{xavi,schulz}; this estimate includes an envelope that considers the cost of hardware, procurement, operations, networking and energy.
Therefore, the estimated origination cost of the ATLAS Open Event Generation amounts to circa CHF $185\textrm{k}$, where each CPU-core is accompanied by 3~GB of memory. The cost of storage can be estimated similarly to compute capacity. Complementing the quote in Figure~\ref{fig:STORAGE-COST}, the total cost of 1~PB of usable disk space is approximately CHF $100\textrm{k}$, which brings the total cost of ATLAS Open Event generation to about CHF $275\textrm{k}$. 

This estimate corresponds to the scenario in which the hardware necessary to produce and store the events is procured and operated \emph{ex novo} on the CERN Tier-0 premises, and the data persist for five years after the event generation.

\subsubsection{Google Cloud impact}
In the same way, it is possible to estimate the financial impact of generating the same data using commercial cloud solutions.
Adopting Google Cloud's C4 standard machines,\footnote{The reference region used on the Google Cloud platform for this and the following estimates is ``europe-west6''.} including 288 CPU cores and 1080~GB of memory (i.e.~3.75~GB per CPU core), and assuming utilisation for 3 years (with the corresponding discount), the cost is USD 7.4 per hour~\cite{google_cloud_compute}. Hence, the previously mentioned 3500 CPU cores of continuous utilisation over two years equals USD 1.58M in costs.

The storage cost $C$ for linear filling from 0 to $S = 900$~TB (or 900000 GB) over $T = 2$ years expressed in hours at rate $r = 0.000027\,\textrm{USD/GB-hour}$, using Google's ``Cloud Storage'' technology~\cite{google_cloud_storage} is:
$$C=r\cdot\frac{ST}{2}\simeq \textrm{USD}\,213\textrm{k}.$$
However, the less realistic scenario where 900~TB are allocated in bulk for two years is more consistent with the assumptions of the CERN Tier-0 estimates, and would correspond to approximately USD 425k in storage costs.
The cost of keeping the produced data on Google Cloud's premises for five years, assuming the same constant rate for storage, would then amount to roughly USD 1.06M.
In addition, the cost of egress to transfer the produced data, according to Google's rates~\cite{google_cloud_network}, would be approximately USD 45 per TB per month, or USD 40k for a one-time transfer.

The total cost of producing the ATLAS Open Event generation campaign over two years and storing it for an additional five years on Google Cloud's premises would then total USD 3.07M, corresponding to CHF 2.42M with the current 0.79~CHF/USD conversion rate~\cite{dollar2CHF}. This estimate excludes the cost of egress, and assumes that no \emph{ad hoc} agreement is stipulated between the collaboration producing the data, and Google Cloud.

Recent studies performed by the ATLAS Collaboration on using Google Cloud resources, where a ``Google Cloud Service Agreement for Public Sector contract'' was negotiated, include an average of 7000 compute cores together with up to 7~PB of storage, and an estimated egress of no more than 0.7~PB per month at a flat rate of USD 56.6k per month~\cite{atlas-gc-2025}.
At the time of writing, the cost of 1000 CPU cores and of 1~PB of storage are approximately equal. Assuming that this statement also holds for the above Public Sector contract, 
this would correspond to about USD 430k of the total USD 1.36M contract value to support 3500 compute cores running at saturation for two years and 900 TB of storage.
This is roughly \SI{80}{\percent} cheaper than requesting resources as a private organisation, but still about \SI{85}{\percent} more expensive than the CERN Tier-0 scenario, without accounting for monthly egress costs.

\begin{table}[ht]
\centering
\begin{tabular}{lcccc}
\toprule
 \textbf{Scenario} & \multicolumn{4}{c}{\textbf{Costs [kCHF]}} \\
 \toprule
 & Compute & Storage & Egress & Persistence (5 years) \\
\midrule
CERN Tier 0 & 185 & 90 & -- & (incl.) \\
Google Cloud & 1245 & 336 & 0.036/TB-month & 841 \\
\midrule
Google Public Sector & 270 & 69 & (incl.) & 173 \\
\bottomrule
\bottomrule
\end{tabular}
\caption{Cost summary for the replication of ATLAS Open Event Generation (900\,TB produced over two years) and storage for five years. Costs expressed in thousands of CHF (kCHF).}
\label{tab:cost-summary}
\end{table}

\subsubsection{Resource Outlook}

HL-LHC projections for ATLAS~\cite{CERN-LHCC-2022-005} foresee \SI{15}{\percent}--\SI{20}{\percent} of the available CPU resources being spent on event generation, and a 2.5--5-fold increase in computing resources in Run~4. By the end of Run~5, an additional factor of 2 increase is expected. Similarly, CMS plans to ramp up event generation to 125B (160B) events/year in Run~4 (Run~5)~\cite{CMS2026}.
Costing this 5--10-fold increase in compute requires a projection for the future cost of hardware, and that of energy.

Early 2026 forecasts project the global semiconductor market to reach about USD $1$ trillion in annual sales~\cite{wsts}, driven by AI demand for logic and memory. Amazon, Alphabet, Meta, and Microsoft plan USD $665$ billion in 2026 AI infrastructure capital expenditures~\cite{amz,google,meta,msft}, over two-thirds of a year's semiconductor output if substantially allocated to hardware. Further, as countries struggle to reconcile growing global energy demands with grid decarbonisation pledges, it is likely that energy prices will rise (c.f.~Section~4 of Ref.~\cite{Banerjee:2023avd}).

Within the Worldwide LHC Computing Grid's (WLCG) flat budget envelope, this market concentration and volatility, including recent RAM price surges, further incentivises efficient, shared event generation to minimise both operational costs and hardware procurement pressure.

\subsection{Environmental impacts}

Although e.g.\ HS23s are a good proxy for operational cost, geographical differences in grid carbon intensity, and energy efficiency and utilisation of computing infrastructure can result in their scaling very differently from environmental costs.   
These environmental costs span the entire life cycle of computing hardware and infrastructure, from its production and transport to point-of-sale, through energy and water consumption in its use phase, and finally recycling and disposal at the end of its useful life, and is quantified in a life cycle assessment (LCA).  A more comprehensive discussion of the environmental impacts of computing in HEP and related disciplines can be found in Ref.~\cite{Banerjee:2023avd}.  

Our focus here will be restricted to the greenhouse gas (GHG) emissions of computing, in this case the MC simulations producing event generation data.  These emissions can be divided into two main parts: `embodied' GHGs due to the production of the computing hardware and housing infrastructure (such as physical data centres), and energy use in operation, including e.g.~cooling costs.
Hardware production emissions are dominated by the manufacture of microchips, particularly the energy-intensive purification of silicon into electrical-grade monocrystalline wafers (a partial LCA for which can be found in Ref.~\cite{Banerjee:2023avd}), and subsequent etching and cleaning, and increases with hardware capability.  It is `paid for' at equipment purchase, and amortised over the hardware lifetime.  Although relevant and up-to-date emissions data can be difficult to find, particularly for products in competitive markets, and produced using proprietary techniques, hardware manufacturers are increasingly providing estimates of production emissions for their products. However the information available lacks sufficient detail for independent verification to be possible.

Embodied carbon can dominate over use-phase emissions for equipment which is underutilised, e.g.\ personal computing equipment.  By contrast, for industrial-scale computing equipment under average conditions, it is the use-phase that dominates emissions~\cite{GuptaChasingCarbon}.\footnote{Progressive grid decarbonisation and increasing server efficiencies will make embodied emissions come to dominate all computing emissions in the future.}  

Use-phase emissions scale roughly as:
\[
\textrm{Compute [CPUs]}\times\textrm{Hardware power [kWh/CPUs]}\times\textrm{grid carbon intensity [gCO$_{2\textrm{e}}$/kWh]}\times\textrm{PUE}\;. \]

\noindent
The PUE, or Power Usage Effectiveness, is a factor that accounts for the `overhead' energy usage of the data centre housing the computing hardware, including energy used in cooling, power delivery and networking.  This can range from 1.1--1.8+, depending on the age and design of the data centre.

The use-phase GHG emissions per event can be computed from the average computational cost of 232~HS23s per event as follows.  A core with an efficiency of 15~HS23s/CPUs converts this to a wall-clock time of approximately 15.5~CPU-seconds, consuming $15.5\;\textrm{s} \times 10\;\textrm{W} = 155\;\textrm{J}$ of energy, assuming 10~W energy usage per core (including processing, RAM, storage and peripherals)\footnote{As a benchmark, Dell's Enterprise Infrastructure Planning Tool~\cite{DellEIPT} quotes a system input power of around 700~W for a Dell PowerEdge server of 2x32 Intel Xeon Platinum cores with 512~GB RAM, resulting in just over 10~W/core.}.  Multiplying by a PUE of 1.45~\cite{WLCG_PUE} and a grid carbon intensity of 245~gCO$_{2\textrm{e}}$/kWh, computed using the grid carbon intensities~\cite{OWID} of WLCG countries, weighted by the fraction of ATLAS event generation that was run in those countries in 2024 and 2025, yields $155\;\textrm{J} \times 1.45 \times 245\;\textrm{gCO}_{2\textrm{e}}\textrm{/kWh}\;/\;(3.6\times10^6\;\textrm{J/kWh}) \approx 0.015\;\textrm{gCO}_{2\textrm{e}}$ emitted per simulated event.

More generally, the per-event use-phase emissions can be expressed as the following scaling relation, where each bracketed factor is a dimensionless ratio equal to unity at the reference values given; substituting a different value for any parameter shows its effect on the result:

\begin{equation}
0.015\,\textrm{gCO$_{2\textrm{e}}$}\left(\frac{\textrm{Efficiency}}{15\frac{\textrm{HS23s}}{\textrm{CPUs}}}\right)^{-1} \left(\frac{\textrm{Power/CPU}}{\SI{10}{W}}\right)\left(\frac{\textrm{PUE}}{1.45}\right)\left(\frac{\textrm{grid carbon intensity}}{245\,\textrm{gCO$_{2\textrm{e}}$/kWh}}\right) \,.
\end{equation}

\noindent
Similarly, using as a benchmark the manufacturer-reported GHG footprint of a Dell 64-core PowerEdge C6520 server of $10.3\pm5.4$ tCO$_{2\textrm{e}}$~\cite{DellPCF}, \SI{16}{\percent} of which is due to manufacturing and transport of hardware, and allocating the resulting embodied emissions of 1648~kgCO$_{2\textrm{e}}$ over a server's estimated 4-year lifespan at full utilisation, yields an embodied emissions per event of

\begin{equation}
0.0033\,\textrm{gCO$_{2\textrm{e}}$}\left(\frac{\textrm{Efficiency}}{15\frac{\textrm{HS23s}}{\textrm{CPUs}}}\right)^{-1}\left(\frac{\textrm{\# cores}}{64}\right)^{-1}\left(\frac{\textrm{embodied emissions}}{1648\,\textrm{kgCO$_{2\textrm{e}}$}}\right)\left(\frac{4\textrm{ years}}{\textrm{lifespan}}\right)\,,
\end{equation}

\noindent assuming \SI{100}{\percent} CPU saturation and a \SI{5}{\percent} failure rate.  This results in a total GHG footprint estimate for ATLAS Open Event Generation of:
\[
195\,\textrm{tCO}_{2\textrm{e}}\,\textrm{(use phase)} + 42\,\textrm{tCO}_{2\textrm{e}}\,\textrm{(embodied)} \,,
\]
obtained by multiplying the per-event emissions by $12.8\times10^{9}$ events.
ATLAS produces an order of magnitude more event generation data annually~\cite{ATLAS:2024vdo}.  Although CMS does not release annual breakdowns of compute usage for MC event generation, their total Run 2 (2015-2018) simulated dataset is comparable in size~\cite{CMS2026}. Assuming substantial overlap between these simulated datasets, centralising ATLAS and CMS event generation has the potential to save up to 2400\,tCO$_{2\textrm{e}}$ in each year, equivalent to the emissions due to over a thousand transatlantic return flights (for a single passenger and average flight occupancy).

Using commercial cloud resources for event generation would result in  a substantially smaller operational footprint, owing to the use of renewable electricity sources at large data centres~\cite{GoogleGridCarbonData}, and more modern data centre infrastructure, with an average PUE of 1.09 \cite{GoogleSustainabilityReport}. For instance, the grid carbon intensity of Google's `europe-west-6' data centre is 15.05 gCO$_{2\textrm{e}}/$kWh \cite{GoogleGridCarbonData}, a factor of 15 below the WLCG average. However, the embodied footprint of these data centres is anticipated to be larger than that for the WLCG sites, owing to more frequent hardware turnover, and correspondingly shorter amortisation timelines.

SM event generation is also replicated independently by the phenomenology and tools communities on a needs basis for individual studies including recasting and proposed new searches, albeit at a much smaller scale than by the experimental collaborations.  We can estimate the number of such studies annually by extracting from the arXiv the number of papers referencing MC event generators, around 150 in 2025, and assuming each paper requires $\mathcal{O}(3\text{M})$ showered and hadronised MC events,\footnote{Taking as a benchmark a 13 TeV BSM search in the $e\mu$ channel, with dominant backgrounds SM $t\bar{t}$ and single top production, with the $W$ bosons decaying to $e^\pm\mu^\mp$, with cross sections 18.9 pb and 1.6 pb, respectively, we obtain around 2.8M events for an integrated luminosity of 139~fb$^{-1}$.} and scale by a `pragmatic factor' of 3 to account for repeated generation due to errors.  Simulations for high luminosity or for larger centre-of-mass energy, or for multijet backgrounds, would require significantly more generated events.  If we used the same server and grid specifications as above,  we would obtain an annual GHG footprint for event generation of around $16\,\textrm{tCO}_{2\textrm{e}}\,\textrm{(use phase)} + 4\,\textrm{tCO}_{2\textrm{e}}\,\textrm{(embodied)}$.  This number should be taken with a grain of salt, however, as it has a large sensitivity to the carbon intensity of the grid that powers the hardware used for the generation, which can vary by over an order of magnitude, from 30 gCO$_{2e}$, to over 600 gCO$_{2e}$ within WLCG countries~\cite{OWID}.  It also depends to a lesser extent on the hardware specifications; using the numbers corresponding to a modern laptop would slightly decrease the environmental footprint of phenomenology studies due to less power-hungry processors and a longer amortised lifetime.

\subsubsection{Resource Outlook}
Global data centre electricity consumption is projected to more than double by 2030, driven primarily by AI workloads~\cite{IEA2025}. This surge risks slowing the decarbonisation of electricity grids, particularly in regions hosting major computing infrastructure, while increasing competition for renewable energy procurement. In parallel, progressive grid decarbonisation in several WLCG member states~\cite{OWID} is expected to gradually reduce use-phase emissions per unit of compute, making embodied emissions an increasingly dominant component of the total GHG footprint.

The 5--10-fold increase in computing resources projected for LHC Runs~4 and ~5~\cite{CERN-LHCC-2022-005,CMS2026}, would scale the environmental footprint of event generation proportionally, unless countered by measures such as shared generation.

%%%%%%%%%%%%%%%%%%%%%%%%%%%%%%%%%%%%%%%%
%%%%%%%%%%%%%%%%%%%%%%%%%%%%%%%%%%%%%%%%
\section{Future Directions} 
\label{sec:future}
%%%%%%%%%%%%%%%%%%%%%%%%%%%%%%%%%%%%%%%%
%%%%%%%%%%%%%%%%%%%%%%%%%%%%%%%%%%%%%%%%

\subsection{Sharing of parton-level events}
\label{sec:partonlevel}

In addition to particle-level event samples, there is a strong case for future community releases of \textit{parton-level} event samples,
for example in plaintext \lhe format~\cite{Alwall:2006yp,Andersen:2014efa}
or in more modern \hdffive{}-based \lhehfive format~\cite{Hoche:2019flt,Bothmann:2023ozs}.
In contrast to particle-level samples --
which include the simulation of bremsstrahlung via parton showers,
as well as non-perturbative effects such as MPI and hadronisation -- parton-level samples encode only the fixed-order hard scattering and its associated metadata, including the PDFs used for the incoming hadrons.

Fixed-order matrix-element generators such as \powheg and \madgraph are explicitly designed to produce parton-level event samples, typically written in the \lhe{}-based formats. These files encode the hard-scattering matrix elements and event-by-event information needed for subsequent showering. In contrast, general-purpose event generators that include parton showers -- such as \sherpa, \pythia or \herwig{} -- most commonly write out particle-level events, typically in the \hepmc format, as their primary output. While these tools may internally generate or consume parton-level events as part of matching or merging workflows, the hard-scattering step is usually treated as an internal component of the simulation chain rather than as a persistently stored product. Dedicated parton-level generator codes targeting GPU-accelerated hard-scattering calculations, such as \pepper~\cite{Bothmann:2023gew} and the CUDACPP plug-in~\cite{Hagebock:2025jyk} for \madgraph (which will be renamed to \textsc{MadMatrix} in the upcoming \textsc{MadGraph7} release), focus exclusively on the efficient production of parton-level event samples, with these samples forming their primary and persistent output. 

While parton-level event samples are substantially more compact than their particle-level counterparts, this should not be interpreted as a direct reduction in overall storage requirements for experimental production workflows. In realistic use cases, particle-level event samples will still need to be produced and archived, and the release of parton-level inputs therefore represents an \textit{additional} data product rather than a replacement.

The key point is instead that the storage footprint of parton-level samples is negligible compared to that of particle-level event records. As a representative example, consider an inclusive multijet-merged sample for Drell-Yan production in association with up to five jets, with the zero-, one-, and two-jet multiplicities calculated at NLO accuracy and generated with \sherpa at \SI{13}{\TeV}.
A compressed \hepmc~3 particle-level sample containing 1000 events occupies
\SI{45}{\mega\byte} (\SI{50}{\mega\byte} when including 150 variation weights to track scale and PDF variations),
whereas the corresponding compressed \lhehfive files
with all information required to fully regenerate this sample (including potential variation weights)
amount to only about \SI{240}{\kilo\byte} in total,
reflecting the much smaller number of particles present at parton level. Against the scale of particle-level storage already required for HL-LHC production campaigns, this additional cost is essentially negligible.

This asymmetry in storage cost enables a qualitatively different workflow: the most computationally expensive stage of event generation -- the fixed-order hard-scattering calculation -- can be produced once and subsequently reused. For recent ATLAS \sherpa setups for $Z$-boson plus jets and top--antitop plus jets production, between 60\,\% and 80\,\% of the total event-generation wall-clock time is spent on generating parton-level events~\cite{Bothmann:2022thx}. Reusing pre-generated parton-level samples therefore has the potential to accelerate full particle-level production campaigns by factors of three to five.

In the HL-LHC era, where both computing and storage resources are under sustained pressure, this makes it feasible to \textit{upgrade} baseline particle-level predictions without having to redo the parton-level generation. Such upgrades may be motivated by improved parton showers, updated non-perturbative models or tunes, or revised theoretical assumptions, and would often be impractical if the hard-scattering step had to be regenerated for each variation. Indeed, for exactly this reason the experiments often internally store parton-level events already.

When shared parton-level event samples are available, experiments and other users can reuse the expensive hard-scattering calculation while retaining complete freedom over the subsequent showering and hadronisation stages. Generating multiple particle-level variants of the same parton-level sample -- using different shower algorithms, hadronisation models, or MPI tunes -- becomes computationally tractable, since the dominant cost has already been amortised. This flexibility is particularly valuable both for experimental systematic studies and for collider phenomenology, where controlled comparisons of modelling choices are essential.

Large parton-level event samples are also of growing importance for machine-learning applications. These include the training of matrix-element surrogate models~\cite{Bishara:2019iwh,Badger:2020uow,Aylett-Bullock:2021hmo,Maitre:2021uaa,Danziger:2021eeg,Winterhalder:2021ngy,Badger:2022hwf,Janssen:2023ahv,Maitre:2023dqz,Bahl:2024gyt,Brehmer:2024yqw,Breso-Pla:2024pda,Bahl:2025xvx,Herrmann:2025nnz,Favaro:2025pgz,Villadamigo:2025our,Beccatini:2025tpk,Bahl:2026qaf}, as well as generative approaches -- such as normalising flows -- used to efficiently sample high-dimensional phase spaces~\cite{Bendavid:2017zhk,Klimek:2018mza,Chen:2020nfb,Bothmann:2020ywa,Gao:2020zvv,Gao:2020vdv,Heimel:2022wyj,Verheyen:2022tov,Heimel:2023ngj,Deutschmann:2024lml,Heimel:2024wph,Bothmann:2025lwg,Janssen:2025zke,Bothmann:2026dar,Heimel:2026hgp,DeCrescenzo:2026tsp}. In such contexts, the ability to share and reuse large, well-documented parton-level datasets is particularly valuable.

The principal challenge associated with sharing parton-level event samples lies in the need for careful alignment between the parton-level generator configuration and the subsequent showering and hadronisation setup. Mismatches in scale choices, recoil schemes, colour or flavour information, or PDF assumptions can result in physically inconsistent samples, potentially wasting both human effort and computing resources. Addressing this risk will require continued community work on standards for passing information from parton-level generators to parton-shower programs, as well as on robust validation and sanity-checking mechanisms that can flag problematic configurations early.

Encouragingly, storing and reusing parton-level samples (e.g., generated with \powheg and \madgraph and then further processed with \pythia or \herwig) is already common practice within the experimental collaborations, and existing examples of sharing LHE files between collaborations, as discussed in Section ~\ref{sec:sharing}, demonstrate success in overcoming the above challenges.

For \sherpa, a traditionally ``end-to-end'' framework including both parton- and particle-level simulation steps, reading and writing parton-level events is a capability that has been added only recently. 
In particular, a recent demonstration has shown that parton-level \lhehfive event samples generated with \sherpa and \pepper can be successfully used as input for particle-level simulations with both \sherpa and \pythia, illustrating the feasibility of combining these different parton-level and particle-level tools within a coherent workflow~\cite{Bothmann:2023ozs}.

Further standardisation and documentation of such pipelines would significantly strengthen the case for routine community releases of parton-level event samples.

\subsection{Accessibility through \hepdata}
\label{sec:HEPData}

Most LHC analyses have an associated record in the \hepdata repository~\cite{HEPData}, providing detailed numerical material.  
These records provide a natural mechanism for linking published analyses to the datasets on which they are based. 
We propose that, whenever the relevant Open Event Generation datasets are publicly available, \hepdata records include DOIs for the corresponding CERN Open Data Portal records, providing a direct mapping from a published result to the (generator-level) samples used to produce it. Where datasets are not yet publicly available, providing a unique internal dataset identifier would still allow community members to understand the precise generator configurations used, and to make targeted requests for public release.

Such linking would benefit reproducibility of LHC analyses, both within and outside the experimental collaborations. A user wishing to reproduce a published ATLAS or CMS result would be able to identify, from the \hepdata record, exactly which Open Event Generation samples were used, retrieve those samples from the CERN Open Data Portal using their DOIs, and reproduce the relevant distributions without needing to independently generate or identify equivalent samples. This workflow is already partially demonstrated by the study described in Section~\ref{sec:ReuseMeasurements}, where ATLAS \sherpa dilepton samples were used to reproduce and extend a CMS Drell-Yan measurement.

In practice, implementing this linking will require a modest extension to the \hepdata submission workflow: authors would be asked, at the time of submission, to provide DOIs or dataset identifiers for the Open Event Generation samples used in their analysis.\footnote{If the internally-used datasets aren't available publicly in their entirety, or slightly different versions were used internally and released publicly, the \hepdata record should note these discrepancies clearly.} This is analogous to the existing expectation that authors provide links to analysis code and statistical models. The \hepdata team has expressed openness to supporting such extensions, and the infrastructure for DOI resolution is already in place through the CERN Open Data Portal. The ATLAS \texttt{atlasopenmagic} tool~\cite{atom25} already provides programmatic access to dataset metadata and DOIs, which could facilitate automated linking for ATLAS samples.

\subsection{Central analysis facilities}
\label{sec:CAF}

As datasets continue to grow in size, and the computational power required for their analysis increases, users’ needs increasingly converge towards the challenges that characterise all distributed computing systems:
\begin{enumerate}
\item performing computation close to where the data is (near-data computing);
\item accessing sufficient computing capacity to complete workloads within a reasonable time frame.
\end{enumerate}
Users require a central facility where they can analyse locally available data, while also having access to the necessary computing resources and curated software stacks to perform a complete physics analysis, from data discovery to analysis execution and results storage.
The purpose of such a central analysis facility is to shield researchers from technical overheads such as resource setup, data access configuration, and software version management.

In recent years, several initiatives have been developed to address these challenges. Platforms such as the ESCAPE VRE and SWAN at CERN integrate data access, compute capabilities, and user management, with the aim of democratising access to scientific computing facilities~\cite{Guerrieri2025}.
Through technologies such as Rucio for data management, and REANA~\cite{reana} for reproducibility and workflow orchestration, these platforms enable users to discover, access, and analyse open or shared datasets seamlessly.

In addition, frameworks such as ServiceX~\cite{servicex} enable near-data processing, where users provide a dataset identifier and a selection statement, and the tools returns the outputs without the need to move data. For use-cases that can fit the pattern ServiceX expects, this provides some of the same benefits as a full-scale analysis facility like remote, rapid processing of the data, efficient scaling, and the streaming of much smaller data volumes.

Finally, since the beginning of 2024, the European Open Science Cloud (EOSC) has launched a campaign to build a network of interconnected computing nodes that can collaboratively share and manage scientific data, knowledge, and resources across thematic and geographical research communities.
As CERN participates in the early build-up phase of this initiative~\cite{eoscfederation}, there is a significant opportunity to support and strengthen the federation by developing and validating use cases, such as those based on Open Event Generation, and by contributing to the requirements to guide and test the evolving infrastructure.

Going forward, three priorities require community attention:
\begin{itemize}
\item Adoption of standard workflow management frameworks, which define analyses as pipelines of dependent tasks and enable reproducibility, provenance tracking, and scalable execution. These frameworks broadly fall into two paradigms: declarative tools such as Snakemake~\cite{snakemake}, where the pipeline is described through configuration and rules, and code-based tools such as Luigi~\cite{luigi} and Dask~\cite{dask}, where workflows are expressed programmatically. WLCG communities are increasingly exploring these pipeline-based architectures to foster a unified user base and address the analysis tools interoperability.
\item Implementation of visual tools or dashboards to build, test and monitor the progress of complex workflows without relying only on terminal-based logging. REANA could be a valid testing ground for this feature.
\item Enhancement of the CERN Open Data Portal to ensure compatibility with tools designed to apply pre-computed fully differential results in the derivation of observables at higher perturbative orders, such as HighTEA~\cite{hightea}, as well as interpolation grid tools, which enable the recalculation of fixed-order predictions, without repeating the full event generation~\cite{fastnlo,applgrid,pineappl}.
\end{itemize}

Addressing these principles and integrating the necessary functionality into the available and upcoming computing platforms would help unlock significant resources for the broader community and ensure that the available Open Data are used to their full potential. 

%%%%%%%%%%%%%%%%%%%%%%%%%%%%%%%%%%%%%%%%%%%%%%%%%%%%%%%%%%%%%%%%%%%%%%%%%%%%%%%%
\section{Conclusions}
\label{sec:conclusion}
%%%%%%%%%%%%%%%%%%%%%%%%%%%%%%%%%%%%%%%%%%%%%%%%%%%%%%%%%%%%%%%%%%%%%%%%%%%%%%%%

We reviewed ongoing efforts in Open Event Generation for the LHC and their uptake by the community. As of present, 
generator-level MC data is provided by the ATLAS and CMS collaborations through the CERN Open Data Portal. 
Our overview of the current practice addressed key aspects including data formats, curation, metadata provision, lessons learned, and directions for future releases. 

We argued that Open Event Generation helps reduce duplication of effort and resource
consumption, and benefits the whole HEP community. To strengthen this claim and illustrate the uptake by the community, we provided concrete use cases and discussed user experiences. Moreover, we discussed the financial and environmental impacts, highlighting the significant benefits of data sharing across a variety of metrics.

This analysis of current generator-level Open Data offerings also brought to light clear opportunities for improvement. The concrete actions we propose are summarised below. It is important to note that the responsibility for these developments need not rest solely on the experimental collaborations; many could equally be realised through community contributions.

An immediate obstacle to improving and extending Open Data offerings is the issue of resources: the hardware and infrastructure housing the data, and support personnel, remain critically limited. 
It is our hope that, with sufficient engagement from the user community and a firm commitment from a major computing resource provider like CERN, it will be possible to allocate sufficient resources to overcome these limitations. 

\textbf{In summary}, Open Event Generation offers a number of compelling advantages for the experimental, phenomenology, and theory communities. 
Raising awareness of and maintaining available offerings and the resources and tools necessary to use them (including effective curation and replacement of any incorrect datasets)  is key to fully realising these advantages. In the long term, expanding Open Data offerings and infrastructure, and establishing a centralised, community-driven LHC MC production group\footnote{The recently established ``Data Sharing and New Workflows'' task force of the LHC MC WG might be a first step in this direction.} present opportunities for both resource savings and improved engagement of the theory and phenomenology communities in the ongoing experimental programme.
Realising these opportunities requires that these initiatives be embedded in the strategy for any future collider program, with the support of the host laboratory, just as safeguarding the Open Data themselves even beyond the lifetime of the experiments is the responsibility of the host laboratory.

%\clearpage
%%%%%%%%%%%%%%%%%%%%%%%%%%%%%%%%%%%%%%%%%
\subsection*{Summary of action items}
%%%%%%%%%%%%%%%%%%%%%%%%%%%%%%%%%%%%%%%%%

\begin{enumerate}

{\bf \item{Short-term actions:}}

\begin{enumerate}
\item Provide stand-alone Open Event Generation tutorials from existing ATLAS and CMS materials, with detailed explanations of metadata, including keywords, and examples that require downloading all files in a dataset (e.g.~including features like a progress bar, error messages and handling, and automatic retries).
\item Develop a conversion tool from CMS formats to \hepmc for cross-compatibility, and compatibility with downstream public analysis tools.
\item Develop capabilities within phenomenology tools like \checkmate or \madanalysis to import generator-level samples through the Open Data Portal, and to access files over a network connection rather than local download, in order to make processing of Open Event Generation samples as straightforward as possible.
\item Expand and harmonise Open Event Generation metadata. Within ATLAS, this could include further documentation of ``physics short'' names and better metadata describing dataset sizes, for example. Across ATLAS and CMS, this could include the harmonisation of available metadata content and format, ensuring accessibility through a single tool if possible (e.g.~providing all metadata via the CERN Open Data Portal, or via a separate python client tool) that is shared across the community and provides a consistent look and feel for all samples.
\item Ensure that all Open Data is available through the CERN Open Data Portal (in particular for ATLAS samples). Where possible, ensure that formats released through the Open Data Portal support streaming (e.g.~using gzipped, un-tarred files for text-based data formats).
\item Understand the substantial size difference between \hepmc-2 and \hepmc-3 formats, and identify any opportunities for space savings in the newer format, including attributes that might not be needed. This is important before the \hepmc-3 format can be used for extended data releases.
\end{enumerate}

{\bf \item{Mid--term actions:}}

\begin{enumerate}
\item Engage with the ALICE and LHCb Collaborations to motivate the release of samples that are of particular interest and importance to the physics programmes of those experiments.  
\item Associate the datasets used in analyses and those released as Open Data via the HEPData repository. Automatic linking of datasets would provide significant documentation benefits to the community.
\item Release a wider variety of BSM event samples, including e.g.~signal samples used in published BSM searches. 
\item Optimize and harmonise data formats and compression for size and ease of access, prioritising standard formats (\hepmc and LHE) over experiment-specific ones. This may necessitate a re-release of existing data, or might be provided only for future sample releases.
\item Develop a central analysis facility, offering computational resources for near-data studies using the Open Event Generation data, in line with what was described in Section~\ref{sec:CAF}.
\item Release parton-level (\lhe) events, at least for SM samples that require com\-pu\-ta\-tion\-ally-intensive high-precision matrix-element calculations, to allow for quick upgrades of baseline particle-level simulations, and for machine-learning applications. 
\end{enumerate}

{\bf \item{Long-term action:}}

\begin{quote}
Centralise the production of MC event samples to a dedicated core LHC team, working in close consultation with MC generator experts and the experimental collaborations. Such a team would be responsible for agreeing on and maintaining a common set of generator configurations, starting with the most important SM processes, producing samples at sufficient statistics for the full HL-LHC programme, and releasing them through the CERN Open Data Portal in an agreed format accessible to all.
\end{quote}

%% alternative proposal by Chris Gutschow:
%\begin{quote}
% Establish a coordinated framework for the production and sharing of reusable MC event samples, working in close consultation with MC generator experts and the experimental collaborations. Such a framework would facilitate agreement on common baseline generator configurations, beginning with the most important SM processes, support the production and release of reusable samples through the CERN Open Data Portal in agreed formats, and enable their reuse across the LHC experiments and the wider community. 
%\end{quote}

    Ideally this would involve:
    \begin{itemize}
        \item an agreement between ATLAS and CMS on a mutually compatible baseline set of generator settings (building on precedents such as in Ref.~\cite{ATL-PHYS-PUB-2023-016});
        \item a clear model for computing resource allocation, for example through a dedicated WLCG allocation or a CERN-hosted production facility; 
        \item a commitment by CERN, alone or in consortium with other major physics labs and research infrastructure providers, to provide long-term storage for the samples;
        \item and a forum where the developers of the event generators can be consulted and discuss with the experiments. 
        \end{itemize}
        
We note that the most important aspect here is coordinating the computationally expensive matrix-element calculation and making parton-level samples reusable, while flexibility may be retained in the subsequent showering and non-perturbative modelling. 

The financial and environmental case for such centralisation, avoiding unnecessary duplication of expensive event generation, is made in Section~\ref{sec:sustainability}. Even under conservative assumptions, the savings from eliminating duplicated event generation across experiments and the phenomenology community amount to a significant fraction of the current generation budget, and will grow substantially as luminosity increases through the HL-LHC era. We recommend that this is considered as part of the resource planning for Run 4.

\end{enumerate}

\noindent
Looking beyond the (HL-)LHC, we see no reason why these actions should not be extended to include event generation carried out in future for legacy and planned particle physics experiments.

In closing, we note that the success of these actions, and of Open Event Generation in general, requires significant resources and commitment, currently provided by CERN. Realising its full potential across cost savings, environmental benefits and science outputs, requires that adequate resourcing be treated not as discretionary, but as a fundamental prerequisite, advocated for by the HEP community, and reliably delivered by providing institutions.

%%%%%%%%%%%%%%%%%%%%%%%%%%%%%%%%%%%%%%%%
%%%%%%%%%%%%%%%%%%%%%%%%%%%%%%%%%%%%%%%%
\section*{Acknowledgements}

We thank Xavier Espinal, Domenico Giordano, Andrea Sciabà, Markus Schulz and Natalia Szczepanek of CERN IT for their insights about the cost of computing resources, operations, and power consumption metrics. Moreover, we thank Dominik Hirschb\"uhl, Maximilian Horzela, Kristin Lohwasser, Peter Millington, and James Whitehead for comments on the manuscript. 
Thanks also to Alastair Basden, Sabine Cr\'ep\'e-Renaudin, Ian Fisk, Jacopo Ghiglieri, Maria Girone, Krzysztof Rolbiecki, Ludovic Scyboz and Hubert Simma for helpful conversations.

%%%%%%%%%%%%%%%%%%%%%%%%%%%%%%%%%%%%%%%%
%%%%%%%%%%%%%%%%%%%%%%%%%%%%%%%%%%%%%%%%

\paragraph{Funding information}
The authors of this document acknowledge partial support by the CHIST-ERA
project OpenMAPP under grants ANR-23-CHRO-0006, UKRI EP/Y036360/1, and NCN 2022/04/Y/ST2/00186; an LPCC-funded MCnet studentship;
the DFG under grant no.\ 396021762 – TRR 257: Particle physics phenomenology after the Higgs discovery; 
PLGrid under grant PLG/2025/018203; 
the Research Ireland Awards Grant 21/PATH-S/9475 (MOREHIGGS) under the SFI-IRC Pathway Program; 
The Royal Society (London, UK) in the form of a Royal Society International Exchanges award, grant no.\ IES$\backslash$R1$\backslash$211138; 
the U.S.~National Science Foundation under grant 2513627; and 
the Office of High Energy Physics of the U.S. Department of Energy under contract DE-AC02-05CH11231. 

This work was moreover supported in part by the \href{https://www.hecap.eco/}{Sustainable HECAP+ Initiative}, which acknowledges financial support from the Environmental Sustainability team at the Science and Technology Facilities Council (STFC) part of UK Research and Innovation (UKRI); UKRI, through a UKRI Future Leaders Fellowship [Grant No. MR/V021974/2] and the European Union as part of the Next Generation EU initiative. Materials and opinions presented represent solely those of the authors and do not necessarily reflect the official views of the European Union or of the European Commission.  Neither the European Union nor the European Commission can be considered responsible for them.

%%%%%%%%%%%%%%%%%%%%%%%%%%%%%%%%%%%%%%%%
%%%%%%%%%%%%%%%%%%%%%%%%%%%%%%%%%%%%%%%%
%\bibliography{references.bib}

%%%%%%%%%%%%%%%%%%%%%%%%%%%%%%%%%%%%%%%%
%%%%%%%%%%%%%%%%%%%%%%%%%%%%%%%%%%%%%%%%

\end{document}